\documentclass[12pt]{article}
\usepackage{amsmath}
\usepackage{graphicx}
\usepackage{hyperref} 
\usepackage{cite}

\makeatletter

\@addtoreset{equation}{section}

\allowdisplaybreaks[2]

\newcommand\refr[1]     {ref.\,\cite{#1}}
\newcommand\refrs[1]    {refs.\,\cite{#1}}

\newcommand\eqn[1]     {eq.\,(\ref{#1})}
\newcommand\eqns[2]    {eqs.\,(\ref{#1}) and~(\ref{#2})}
\newcommand\eqnss[2]   {eqs.\,(\ref{#1})--(\ref{#2})}

\newcommand\fig[1]     {figure~{\ref{#1}}}
\newcommand\figs[2]    {figures~\ref{#1}--\ref{#2}}

\newcommand\sect[1]    {section~{\ref{#1}}}

\newcommand\appx[1]     {appendix~\ref{#1}}

\def\beq{\begin{equation}}
\def\eeq{\end{equation}}
\def\bsp#1\esp{\begin{split}#1\end{split}}
\def\bal#1\eal{\begin{align}#1\end{align}}
\newcommand\nt         {\notag}

\newcommand\tsig[2]    {\sigma^{\rm{#1}}_{#2}}
\newcommand\dsig[2]    {\rd\sigma^{{\rm #1}}_{#2}}
\newcommand\dsiga[3]   {\rd\sigma^{{\rm #1,A}_{\scriptscriptstyle #2}}_{#3}}

\newcommand\la         {\langle}
\newcommand\ra         {\rangle}
\newcommand\bra[3]     {\la {\cal M}_{#1}^{#2}#3|}
\newcommand\ket[3]     {|{\cal M}_{#1}^{#2}#3\ra}
\newcommand\Aket[3]    {|{\cal A}_{#1}^{#2}#3\ra}
\newcommand\braket[4]  {\la {\cal M}_{#1}^{#2}#3|{\cal M}_{#1}^{#4}#3\ra}
\newcommand\SME[3]     {|{\cal M}_{#1}^{#2}{#3}|^2}

\newcommand{\rd}       {{\rm{d}}}
\newcommand{\PS}[1]    {\rd\phi_{#1}}
\newcommand{\mom}[1]   {\{p\}^{#1}}

\newcommand{\Y}[2]		{Y_{#1#2,Q}}

\newcommand{\cC}[2]    {{\cal C}_{#1}^{#2}}
\newcommand{\cS}[2]    {{\cal S}_{#1}^{#2}}
\newcommand{\cCS}[3]   {{\cal C}_{#1}^{~}{\cal S}_{#2}^{#3}}
\newcommand{\cSCS}[2]  {{\cal C}\kern-2pt{\cal S}_{#1}^{#2}}

\newcommand{\rC}       {{\rm C}}
\newcommand{\rS}       {{\rm S}}
\newcommand{\rSCS}     {{\rC}\kern-2pt{\rS}}
\newcommand{\IcC}[2]   {{\rC}_{#1}^{#2}}
\newcommand{\IcS}[2]   {{\rS}_{#1}^{#2}}

\newcommand{\IcSCS}[2] {\rC\kern-2pt\rS_{#1}^{#2}}
\newcommand{\bI}       {\bom{I}}

\newcommand{\Li}       {{\rm Li}}
\newcommand{\Log}[1]   {\ln #1}

\newcommand{\CF}       {C_{\rm{F}}}
\newcommand{\CA}       {C_{\rm{A}}}
\newcommand{\TR}       {T_{\rm{R}}}
\newcommand{\Nc}       {N_{\rm{c}}}
\newcommand{\Nf}       {\ensuremath{n_{\rm{f}}}}

\newcommand{\bT}       {\bom{T}}
\newcommand{\bTsq}[1]  {\bom{T}^2_{#1}}
\newcommand\qb         {{\bar q}}

\newcommand{\fl}[1]   {#1}
\newcommand{\fla}[1]   {#1}

\newcommand{\colorfulNNLO}{{CoLoRFulNNLO}}
\newcommand{\eerad}{{\tt EERAD3}}
\def\AP{Altarelli--Parisi } 
\newcommand\bom[1]     {{\mbox{\boldmath $#1$}}}

\newcommand{\ep}       {\epsilon}
\newcommand{\Sep}       {S_\epsilon}
\newcommand{\Fep}       {S_\epsilon^{\MSbar}}

\newcommand{\cA}       {{\cal A}}

\newcommand{\ri}       {{\rm{i}}}
\newcommand\e          {{\rm e}}
\newcommand\Oe[1]      {\ensuremath{\rm O(\ep^{#1})}}
\newcommand\Oa[1]      {\ensuremath{\rm O(\as^{#1})}}

\newcommand\ldot       {\!\cdot\!}

\newcommand\as  	       {\ensuremath{\alpha_{\rm{s}}}}
\newcommand\asR  	       {\ensuremath{\alpha_{\rm{s}}(\mu)}}

\newcommand\EulerGamma	{\ensuremath{\gamma_{\rm{E}}}}
\newcommand\MSbar	  {\ensuremath{\overline{\rm{MS}}}}

\begin{document}

%%%
%%% Title page
%%%
% ========== ========== ========== ========== ==========

\begin{titlepage}
\begin{flushright}
CERN-PH-TH-2016-138, CP3-16-29, NSF-KITP-16-084
\end{flushright}
\renewcommand{\thefootnote}{\arabic{footnote}}
\par \vspace{10mm}

\begin{center}
{\Large \bf Jet production in the \colorfulNNLO\ method:}
\\[0.25cm]
{\Large \bf event shapes in electron-positron collisions}
\end{center}
\par \vspace{2mm}
\begin{center}
{\bf Vittorio Del Duca}$^{(a)}$\footnote{On leave from Istituto Nazionale di Fisica Nucleare, Laboratori Nazionali di Frascati, Italy.}, 
{\bf Claude Duhr}$^{(b,c)}$\footnote{On leave from the ``Fonds National de la Recherche Scientifique'' (FNRS), Belgium.},
{\bf Adam Kardos}$^{(d,e)}$,
{\bf G\'abor Somogyi}$^{(d)}$,\\[0.25em]
{\bf Zolt\'an Sz\H or}$^{(d,e)}$,
{\bf Zolt\'an Tr\'ocs\'anyi}$^{(d,e)}$~~and~~{\bf Zolt\'an Tulip\'ant}$^{(d)}$\\

\vspace{5mm}

$^{(a)}$ Institute for Theoretical Physics, ETH Z\"urich, 8093 Z\"urich, Switzerland

$^{(b)}$ Theoretical Physics Department, CERN, CH-1211 Geneva 23, Switzerland

$^{(c)}$ Center for Cosmology, Particle Physics and Phenomenology (CP3),
Universit\'{e} Catholique de Louvain,
Chemin du Cyclotron 2,
B-1348 Louvain-La-Neuve,
Belgium

$^{(d)}$ University of Debrecen and MTA-DE Particle Physics Research Group 
H-4010 Debrecen, PO Box 105, Hungary

$^{(e)}$ Kavli Institute for Theoretical Physics, University of California, 
Santa Barbara, CA 93106, USA

\vspace{5mm}

\end{center}

\par \vspace{2mm}
\begin{center} {\large \bf Abstract} \end{center}
\begin{quote}
\pretolerance 10000

We present the \colorfulNNLO\ method to compute higher order radiative corrections 
to jet cross sections in perturbative QCD. We apply our method to the computation 
of event shape observables in electron-positron collisions at NNLO accuracy and 
validate our code by comparing our predictions to previous results in the literature. 
We also calculate for the first time jet cone energy fraction at NNLO.

\end{quote}

\vspace*{\fill}
\begin{flushleft}
June 2016

\end{flushleft}
\end{titlepage}

\setcounter{footnote}{0}
\renewcommand{\thefootnote}{\arabic{footnote}}

%%%
%%% Introduction
%%%
% ========== ========== ========== ========== ==========

\section{Introduction}
\label{sec:intro}

The strong coupling $\as$ is one of the most important parameters of the 
standard model. A clean environment for determining $\as$ is the study of event shape 
distributions in electron-positron collisions. Particularly well suited for this task 
are quantities related to three-jet events, as the leading term in a perturbative 
description of such observables is already proportional to the strong coupling. 
Accordingly, three-jet event shapes were measured extensively, especially 
at LEP \cite{Heister:2003aj,Abdallah:2003xz,Achard:2004sv,Abbiendi:2004qz}. 
The precision of experimental measurements calls for an equally precise 
theoretical description of these quantities. Because the strong interactions occur only 
in the final state, non-perturbative QCD corrections are restricted to hadronization 
and power corrections. These corrections can be determined either by extracting them 
from data by comparison to Monte Carlo predictions or by using analytic models. Hence, 
the precision of the theoretical predictions is mostly limited by the truncation of the 
perturbative expansion in the strong coupling. 

Current state-of-the art includes next-to-next-to-leading order (NNLO) predictions for 
the three-jet event shapes of thrust, heavy jet mass, total and wide jet broadening, 
$C$-parameter and the two-to-three jet transition variable $y_{23}$ 
\cite{GehrmannDeRidder:2007hr,Weinzierl:2009ms}, 
as well as for oblateness and energy-energy correlation \cite{DelDuca:2016csb}. 
Next-to-leading order (NLO) predictions for the production of up to five jets 
\cite{Signer:1996bf,Nagy:1997yn,Campbell:1998nn,Weinzierl:1999yf,Frederix:2010ne} 
(and up to seven jets in the leading color approximation \cite{Becker:2011vg}) are also known.
Moreover, logarithmically enhanced contributions to event shapes can be resummed at up to 
next-to-next-to-leading logarithmic (NNLL) accuracy 
\cite{Catani:1992ua,Banfi:2004yd,Banfi:2014sua,Chien:2010kc,Becher:2012qc} 
and even at next-to-next-to-next-to-leading logarithmic ($\rm N^3LL$) accuracy 
for some observables \cite{Abbate:2010xh,Hoang:2014wka}.

In addition to its phenomenological relevance, three-jet production in electron-positron 
collisions is also an ideal testing ground for developing general tools and techniques for 
higher-order calculations in QCD. The straightforward evaluation of radiative corrections 
in QCD is hampered by the presence of infrared singularities in intermediate stages of the 
calculation which cancel in the final physical results for these observables. Nevertheless, 
they must be regularized and their cancellation has to be made explicit before any numerical 
computation can be performed. 
This turns out to be rather involved for fully differential cross sections at NNLO and 
constructing a method to regularize infrared divergences has been an ongoing task for many 
years \cite{
%%%%%%%%%% Sector decomposition: Binoth, Heinrich %%%%%%%%%%
Binoth:2000ps,
Binoth:2004jv,
%%%%%%%%%% Reverse unitarity: Anastasiou %%%%%%%%%%
Anastasiou:2003gr,
%%%%%%%%%% Sector decomposition w. non-linear mappings: Anastasiou %%%%%%%%%%
Anastasiou:2010pw,
%%%%%%%%%% Weinzierl %%%%%%%%%%
Weinzierl:2003fx,
Weinzierl:2003ra,
%%%%%%%%%% qT: Catani %%%%%%%%%%
Catani:2007vq,
%%%%%%%%%% Antenna method: [23] of 1605.04295 %%%%%%%%%%
GehrmannDeRidder:2005cm,
GehrmannDeRidder:2005aw,
GehrmannDeRidder:2005hi,
Daleo:2006xa,
Daleo:2009yj,
Gehrmann:2011wi,
Boughezal:2010mc,
GehrmannDeRidder:2012ja,
Currie:2013vh,
%%%%%%%%%% STRIPPER: Czakon %%%%%%%%%%
Czakon:2010td,
Czakon:2011ve,
Czakon:2014oma,
%%%%%%%%%% STRIPPER: Petriello %%%%%%%%%%
Boughezal:2011jf,
%%%%%%%%%% Our stuff %%%%%%%%%%
Somogyi:2005xz,
Somogyi:2006cz,
Somogyi:2006da,
Somogyi:2006db,
Somogyi:2008fc,
Aglietti:2008fe,
Somogyi:2009ri,
Bolzoni:2009ye,
Bolzoni:2010bt,
DelDuca:2013kw,
Somogyi:2013yk,
%%%%%%%%%% N-jettiness: Petriello %%%%%%%%%%
Boughezal:2015dva,
%%%%%%%%%% N-jettiness: Gaunt %%%%%%%%%%
Gaunt:2015pea}.

In this paper we present a general subtraction scheme to compute fully differential 
predictions at NNLO accuracy, called \colorfulNNLO\ (Completely Local subtRactions for 
Fully differential predictions at NNLO accuracy) \cite{Somogyi:2005xz,Somogyi:2006cz,
Somogyi:2006da,Somogyi:2006db,Somogyi:2008fc,Aglietti:2008fe,Somogyi:2009ri,Bolzoni:2009ye,
Bolzoni:2010bt,DelDuca:2013kw,Somogyi:2013yk}. 
The method uses the known universal factorization properties of QCD matrix elements in 
soft and collinear limits \cite{Campbell:1997hg,Catani:1998nv,DelDuca:1999iql,Berends:1988zn,
Catani:1999ss,Bern:1997sc,Kosower:1999xi,Kosower:1999rx,Bern:1999ry} 
to construct completely local subtraction terms which regularize infrared singularities 
associated with unresolved real emission. Virtual contributions are rendered finite by 
adding back the subtractions after integration and summation over the phase space and 
quantum numbers (color and flavor) of the unresolved emission. We have worked out the 
method completely for processes with a colorless initial state and involving any number 
of colored massless particles in the final state. We validate our method and code by 
computing NNLO corrections to three-jet event shape variables and comparing our predictions 
to those available in the literature \cite{GehrmannDeRidder:2007hr,Weinzierl:2009ms}. 
We also present here for the first time the computation of the jet cone energy fraction 
(JCEF) at NNLO accuracy. We note that the \colorfulNNLO\ method has already been successfully 
applied to compute NNLO corrections to differential distributions describing the decay of a 
Higgs boson into a pair of b-quarks \cite{DelDuca:2015zqa}, as well as to the computation 
of oblateness and energy-energy correlation in $e^+e^-\to 3$ jet production 
\cite{DelDuca:2016csb}.

The paper is structured as follows: after introducing our notation and conventions 
in \sect{sec:notation}, we present the \colorfulNNLO\ method in \sect{sec:colorfulNNLO}. 
In \sect{sec:ee3jets} we describe the application of the general framework to the 
specific case of three-jet production. In particular, we show that the double virtual 
contribution is free of singularities. Our predictions for event shape observables 
follow in \sect{sec:event}. We draw our conclusions and give our outlook in 
\sect{sec:conclusions}.

%%%
%%% Notation and conventions
%%%
% ========== ========== ========== ========== ==========

\section{Notation and conventions}
\label{sec:notation}

%%%
%%% Phase space and kinematics
%%%

\subsection{Phase space and kinematics}
\label{ssect:PS_kin}

The  phase space measure in $d=4-2\ep$ dimensions for a total incoming momentum $Q^\mu$ 
and $n$ massless outgoing particles reads
\beq
\PS{n}(Q^2) \equiv \PS{n}\big(p_1^\mu,\ldots,p_n^\mu;Q^\mu\big) = 
	\Bigg[\prod_{i=1}^{n}\frac{\rd^d p_i}{(2\pi)^{d-1}}
		\delta_{+}(p_i^2)\Bigg]
	(2\pi)^d \delta^{(d)}\big(p_1^\mu + \ldots + p_n^\mu - Q^\mu\big)\,.
\eeq
Throughout the paper, we will use $y_{ik}$ to denote twice the
dot-product of two momenta, scaled by the total momentum squared $Q^2$.
For example,
\beq
y_{ik} = \frac{2p_i\cdot p_k}{Q^2} = \frac{s_{ik}}{Q^2}
\qquad\mbox{and}\qquad
y_{iQ} = \frac{2p_i\cdot Q}{Q^2}\,.
\label{eq:yik-def}
\eeq
We also introduce the combination
\beq
Y_{ik,Q} = \frac{y_{ik}}{y_{iQ} y_{kQ}}
\label{eq:YikQ-def}
\eeq
for later convenience.

%%%
%%% Matrix elements
%%%

\subsection{Matrix elements}
\label{ssect:ME}

We use the color and spin space notation of \refr{Catani:1996vz} where 
the renormalized matrix element for a given process with $n$ particles in the 
final state, $\ket{n}{}{}$, is a vector in color and spin space, normalized 
such that the squared matrix element summed over colors and spins is given by
\beq
\SME{n}{}{} = \braket{n}{}{}{}\,.
\eeq
The renormalized matrix element has the following formal loop expansion
\beq
\ket{n}{}{} = 
	\ket{n}{(0)}{} 
	+ \ket{n}{(1)}{} 
	+ \ket{n}{(2)}{} 
	+ \ldots \,,
\eeq
where the superscript denotes the number of loops.
We will always consider matrix elements computed in conventional dimensional
regularization (CDR) with \MSbar\ subtraction.  We introduce the
following notation to indicate color-correlated squared matrix elements
(obtained by the insertion of color charge operators
between $\bra{n}{(\ell_1)}{}$ and $\ket{n}{(\ell_2)}{}$):
\beq
\bsp
\braket{}{(\ell_1)}{}{(\ell_2)} \otimes \bT_i \ldot \bT_k &=
     \bra{}{(\ell_1)}{} \,\bT_i \ldot \bT_k \, \ket{n}{(\ell_2)}{}\,,
\\[2mm]
\braket{}{(\ell_1)}{}{(\ell_2)}\otimes \{\bT_i \ldot \bT_k, \bT_j \ldot \bT_l\} &=
     \bra{}{(\ell_1)}{} \{\bT_i \ldot \bT_k, \bT_j \ldot \bT_l\} \ket{}{(\ell_2)}{}
\,.
\esp
\label{eq:otimes-def}
\eeq
The color charge algebra for the product 
$\sum_a (\bT_i)^a (\bT_k)^a \equiv \bT_i \ldot \bT_k$ is
\beq
\bT_i \ldot \bT_k = \bT_k \ldot \bT_i\,,\quad\mbox{if}\quad i \ne k
\qquad\mbox{and}\qquad
\bT_i^2 = C_i\,,
\eeq
where $C_i$ is the quadratic Casimir operator in the representation of particle $i$. 
We use the customary normalization $\TR = 1/2$, and so $\CA = 2\TR\Nc = \Nc$ in the adjoint 
and $\CF = \TR(\Nc^2-1)/\Nc = (\Nc^2-1)/(2\Nc)$ in the fundamental representation.

%%%
%%% Ultraviolet renormalization
%%%

\subsection{Ultraviolet renormalization}
\label{ssect:renorm}

In massless QCD renormalized amplitudes $\ket{n}{}{}$ are obtained 
from the corresponding unrenormalized amplitudes $\Aket{n}{}{}$
by replacing the bare coupling $\as^B$ with the dimensionless renormalized 
coupling $\as\equiv\asR$ computed in the \MSbar\ scheme and evaluated at the 
renormalization scale $\mu$,
\beq
\as^B \mu_0^{2\ep}\,\Fep = \as \mu^{2\ep}
	\bigg[1 - \frac{\as}{4\pi}\frac{\beta_0}{\ep}
+ \left(\frac{\as}{4\pi}\right)^2
  \left(\frac{\beta_0^2}{\ep^2} - \frac{\beta_1}{2\ep}\right)
+ \Oa{3}\bigg]\,,
\label{eq:asRenorm}
\eeq
where
\beq
\beta_0 = \frac{11\CA}{3} - \frac{4\Nf \TR}{3}\,,
\quad
\beta_1 =  \frac{34}3\CA^2 - \frac{20}3\CA\TR\Nf - 4\CF\TR\Nf\,,
\label{eq:beta}
\eeq
and $\Fep = (4\pi)^\ep \exp(-\ep \EulerGamma)$ corresponds to \MSbar\
subtraction, with $\EulerGamma=-\Gamma'(1)$ the Euler--Mascheroni constant. 
Although the factor $(4\pi)^\ep \exp(-\ep \EulerGamma)$ is often abbreviated 
as $\Sep$ in the literature, we reserve the latter to denote 
\beq
\Sep = \frac{(4\pi)^\ep}{\Gamma(1-\ep)}\,,
\eeq
which emerges in the integration of the angular part of the phase space
in $d = 4 - 2\ep$ dimensions.
If the loop expansion of the unrenormalized amplitude is written as
\beq
\Aket{m}{}{} = (4\pi \as^B)^{\frac{q}2}\bigg[
  \Aket{m}{(0)}{}
+ \frac{\as^B}{4\pi}\,\Aket{m}{(1)}{}
+ \left(\frac{\as^B}{4\pi}\right)^2\Aket{m}{(2)}{}
+ \Oa{3}\bigg]\,,
\eeq
(with $q=m-2$, where $m$ is the number of massless final-state partons in the Born 
process) then using the substitution in \eqn{eq:asRenorm} the relations between
the renormalized and the unrenormalized  amplitudes are given as follows:
\beq
\ket{m}{(0)}{} = C(\mu,\mu_0,q;\ep) \Aket{m}{(0)}{}
\,,
\label{eq:renA0}
\eeq
\beq
\ket{m}{(1)}{} = C(\mu,\mu_0,q;\ep) \frac{\as}{4 \pi}
\left[\left(\frac{\mu^2}{\mu_0^2}\right)^\ep\left(\Fep\right)^{-1}\Aket{m}{(1)}{}
 - \frac{q}{2}\frac{\beta_0}{\ep}\,\Aket{m}{(0)}{}\right]
\label{eq:renA1}
\eeq
and
\beq
\bsp
\ket{m}{(2)}{} = C(\mu,\mu_0,q;\ep) \left(\frac{\as}{4 \pi}\right)^2
&\Bigg[\left(\frac{\mu^2}{\mu_0^2}\right)^{2\ep}\left(\Fep\right)^{-2}\Aket{m}{(2)}{}
 - \frac{q+2}{2}\frac{\beta_0}{\ep}
   \left(\frac{\mu^2}{\mu_0^2}\right)^\ep\left(\Fep\right)^{-1}\Aket{m}{(1)}{}
\label{eq:renA2}
\\&
 + \frac{q}{2}\left(\frac{q+2}{4}\frac{\beta_0^2}{\ep^2}
  - \frac{\beta_1}{\ep}\right)\,\Aket{m}{(0)}{} \Bigg]\,,
\esp
\eeq
where 
\beq
C(\mu,\mu_0,q;\ep) =
(4\pi \as)^{\frac{q}2}
\left(\frac{\mu^2}{\mu_0^2}\right)^{\frac{q}2 \ep}
\left(\Fep\right)^{-\frac{q}2} \,.
\label{eq:CBorn}
\eeq
The role of the factors of $(\mu^2/\mu_0^2)^{\ep}$ is to change the regularization 
scale to the renormalization scale so that the renormalized amplitudes in 
\eqnss{eq:renA0}{eq:renA2} only depend on $\mu$. Furthermore, after the IR poles
are canceled in a fixed order computation, we may set $\ep = 0$, therefore, the 
factors of $(\mu^2/\mu_0^2)^{\ep}$ and $\Fep$ in $C(\mu,\mu_0,q;\ep)$ do not
give any contribution, so we may perform the replacement
\beq
C(\mu,\mu_0,q;\ep) \to (4\pi \as)^{\frac{q}2}\,.
\eeq

%%%
%%% Jet production in \colorfulNNLO\
%%%
% ========== ========== ========== ========== ==========

\section{Jet production in \colorfulNNLO}
\label{sec:colorfulNNLO}

We consider the production of $m$ jets from a colorless initial state as in, e.g., 
Higgs boson decay or electron-positron annihilation into hadrons. In perturbative 
QCD the cross section for this process is given by an expansion in powers of the 
strong coupling \as. At NNLO accuracy we retain the first three terms in this 
expansion
\beq
\tsig{}{} = \tsig{LO}{} + \tsig{NLO}{} + \tsig{NNLO}{} + \ldots\,.
\label{eq:sigma_tot}
\eeq
The leading order contribution is simply given by the integral of the fully 
differential Born cross section $\dsig{B}{m}$ of $m$ final-state partons over 
the the available $m$-parton phase space defined by the observable $J$, (often 
called jet function)
\beq
\tsig{LO}{}[J] = \int_m \dsig{B}{m} J_m\,.
\label{eq:sigma_LO}
\eeq
Here and in the following $J_m$ denotes the value of the infrared-safe observable $J$ 
evaluated on a final state with $m$ partons.

%%%
%%% The NLO correction
%%%

\subsection{The NLO correction}
\label{sec:colorfulNNLO-NLO}

The NLO correction is a sum of the real radiation and one-loop virtual terms,
\beq
\tsig{NLO}{}[J] = \int_{m+1} \dsig{R}{m+1} J_{m+1} + \int_{m} \dsig{V}{m} J_{m} \,,
\label{eq:sigma_NLO}
\eeq
both divergent in four dimensions. These two contributions can be made finite 
simultaneously by subtracting and adding back a suitably defined approximate 
cross section $\dsiga{R}{1}{m+1}$,
\beq
\tsig{NLO}{}[J] = 
	\int_{m+1} \Big[\dsig{R}{m+1} J_{m+1} - \dsiga{R}{1}{m+1} J_{m}\Big]_{d=4} 
	+ \int_{m} \Big[\dsig{V}{m} + \int_1 \dsiga{R}{1}{m+1}\Big]_{d=4} J_{m}\,.
\label{eq:sigma_LO_reg}
\eeq
Several prescriptions are available for the explicit construction of the approximate 
cross section \cite{Frixione:1995ms,Catani:1996vz,Nagy:1996bz,Somogyi:2006cz,Somogyi:2009ri}. 
Specifically in the \colorfulNNLO\ framework, it is written as
\beq
\dsiga{R}{1}{m+1} = \frac{1}{2s} \PS{m+1}(Q^2) \cA_1 \SME{m+1}{(0)}{}\,,
\eeq
where the approximate matrix element for processes with $m+1$ partons in the 
final state is given by \cite{Somogyi:2006cz,Somogyi:2006da},
\beq
\cA_1 \SME{m+1}{(0)}{} = 
	\sum_{r=1}^{m+1} 
	\Bigg[ \sum_{\substack{i=1 \\ i\ne r}}^{m+1} \frac1{2} \cC{ir}{(0,0)}
	- \Bigg( \cS{r}{(0,0)} 
		- \sum_{\substack{i=1 \\ i\ne r}}^{m+1} \cC{ir}{}\cS{r}{(0,0)} \Bigg) 
\Bigg] \,.
\label{eq:A1def}
\eeq
On the right-hand side of \eqn{eq:A1def}, $\cC{ir}{(0,0)}$ and $\cS{r}{(0,0)}$ denote   
counterterms which regularize the ${\vec p}_i||{\vec p}_r$ collinear limit and 
the $p_r^\mu\to 0$ soft limit in arbitrary dimensions. The role of the 
$\cC{ir}{}\cS{r}{(0,0)}$ soft-collinear counterterm is to make sure that no double 
subtraction takes place in the overlapping soft-collinear phase space region.
These counterterms were all defined explicitly in \refrs{Somogyi:2006cz,Somogyi:2006da}. 
In our convention the indices of $\cC{ir}{(0,0)}$ are not ordered, 
$\cC{ir}{(0,0)} = \cC{ri}{(0,0)}$. As the sums in \eqn{eq:A1def} over $i$ and $r$ 
are likewise not ordered, the factor of $\frac12$ is needed so that we count
each collinear limit precisely once. Finally, the meaning of the superscript 
$(\ell_1,\ell_2)$ is the following: The corresponding counterterm involves the product 
of an $\ell_1$-loop unresolved kernel (an \AP splitting function or a soft eikonal current) 
with an $\ell_2$-loop squared matrix element (in color or spin space). Specifically, 
$(0,0)$ means that in these subtraction terms a tree level collinear or soft function 
acts on a tree level reduced matrix element.  Such superscripts will appear also for 
other counterterms throughout the paper.

Importantly, the approximate matrix element in \eqn{eq:A1def} takes into account 
all color and spin correlations in infrared limits, and hence it is a completely 
local and fully differential regulator of the real emission matrix element over the 
$m+1$-particle phase space. The complete locality of the subtraction is a necessary 
condition for the regularized real contribution,
\beq
\tsig{NLO}{m+1}[J] = 
	\int_{m+1} \Big[\dsig{R}{m+1} J_{m+1} - \dsiga{R}{1}{m+1} J_{m}\Big]_{d=4}\,,
\label{eq:sigNLO_m+1}
\eeq
to be well-defined in four dimensions.
As pointed out long ago \cite{Catani:1996vz}, when the subtraction terms are not 
fully local, for instance because spin correlations in gluon decay are neglected,
the evaluation of the difference 
$\int_{m+1}\left[\dsig{R}{m+1} J_{m+1} - \dsiga{R}{1}{m+1} J_{m}\right]_{d=4}$ usually 
involves double angular integrals of the type $\int_{-1}^{1}\rd(\cos\theta)\int_{0}^{2\pi}
\rd\phi\, \cos\phi/(1-\cos\theta)$ where $\phi$ is the azimuthal angle. These integrals are
 ill-defined. If their numerical integration is attempted, one can obtain 
any answer whatsoever (including the correct one) depending on the details of the integration 
procedure. (The correct answer is obtained by performing the integral analytically before 
going to four dimensions: $\int_{-1}^{1}\rd(\cos\theta)\,\sin^{-2\ep}\theta
\int_{0}^{2\pi}\rd\phi\,\,\sin^{-2\ep}\phi\cos\phi/(1-\cos\theta) = 0$.) Thus non-local 
subtractions alone are not sufficient to define correctly $\tsig{NLO}{m+1}[J]$. 
Rather, the definition must be supplemented by the precise specification of an integration 
procedure which must be shown to give the correct numerical values for all integrals that 
are finite away from $d=4$, but whose four-dimensional value is ill-defined. 
As in \colorfulNNLO\ the subtractions are completely local, \eqn{eq:sigNLO_m+1} 
is well-defined in four dimensions as it is and may be computed with whatever numerical 
procedure is most convenient. These remarks apply also to the regularized double real 
and real-virtual cross sections in \eqns{eq:sigmaNNLOm+2}{eq:sigmaNNLOm+1} which enter 
the NNLO correction.

Turning to the virtual contribution, the Kinoshita--Lee--Nauenberg (KLN) theorem 
ensures that the integral of the approximate cross section precisely cancels the 
divergences of the virtual piece for infrared-safe observables, so adding back what 
we have subtracted from the real correction, the virtual contribution becomes finite 
as well. We have performed the integration of the various subtraction terms analytically 
in \refr{Somogyi:2006cz} and here we only quote the result, which can be written as,
\beq
\int_1\dsiga{R}{1}{m+1} = 
	\dsig{B}{m} \otimes \bI_1^{(0)}(\{p\}_m;\ep)\,,
\label{eq:INTsigmaRdiffA1}
\eeq
where the $\otimes$ product is defined in \eqn{eq:otimes-def} and the
insertion operator is in general given by \cite{Somogyi:2006cz}%
\footnote{The expansion parameter in \refr{Somogyi:2006cz} was chosen
$\as/\Fep$ implicitly, with the harmless factor $1/\Fep$ suppressed.
For the sake of clarity we reinstate the factor $1/\Fep$ here, as well
as in all other insertion operators in eqs.\ (\ref{eq:I11}),
(\ref{eq:I1100}), (\ref{eq:I12}) and (\ref{eq:I2}) below.}
\beq
\bI_1^{(0)}(\{p\}_m;\ep) = 
	\frac{\as}{2\pi}\frac{\Sep}{\Fep}
	\bigg(\frac{\mu^2}{Q^2}\bigg)^\ep
	\sum_{i=1}^m\bigg[\IcC{1,i}{(0)}(y_{iQ};\ep) \bT_i^2 
		+ \sum_{\substack{k=1 \\ k\ne i}}^m 
		\IcS{1}{(0),(i,k)}(Y_{ik,Q};\ep) \bT_i \ldot \bT_k\bigg]\,.
\label{eq:I10}
\eeq
The variables $Q^\mu$, $y_{iQ}$ and $Y_{ik,Q}$ were defined in
\sect{ssect:PS_kin}. The kinematic functions $\IcC{1,i}{(0)}(y_{iQ};\ep)$ and
$\IcS{1}{(0),(i,k)}(Y_{ik,Q};\ep)$ have been computed as Laurent
expansions in $\ep$ in \refr{Somogyi:2006cz}. They are needed up to
finite terms in a computation at NLO accuracy and up to $\Oe{2}$ 
in a computation at NNLO accuracy. We present these kinematic functions
explicitly up to $\Oe{}$ in \appx{appx:I10}, which is sufficient for
checking the cancellation of the $\ep$-poles at NNLO analytically.
We note that there is no one-to-one correspondence between the unintegrated 
subtraction terms in \eqn{eq:A1def} and the kinematic functions that appear 
in \eqn{eq:I10}. The latter are obtained from the former by integrating over 
the unresolved momentum as well as summing over all unobserved quantum numbers 
(color and flavor), and organizing the result in color and flavor space. 
Loosely speaking, the integrated form of $\cC{ir}{(0)}$ enters $\IcC{1,i}{(0)}$
and that of $\cS{r}{(0)}$ enters $\IcS{1}{(0),(i,k)}$. We are, however, free to 
assign the integrated form of $\cC{ir}{}\cS{r}{(0)}$ to either of the 
integrated counterterms. This final organization was performed differently 
in \refr{Somogyi:2006cz} and in this paper. In \refr{Somogyi:2006cz} we grouped the 
integrated form of $\cC{ir}{}\cS{r}{(0)}$ into $\IcS{1}{(0),(i,k)}$, 
while here we find it more convenient to group it into $\IcC{1,i}{(0)}$. 
Before moving on, let us present the universal pole structure of $\bI_1^{(0)}(\{p\}_m;\ep)$ 
for arbitrary number $m$ of final-state partons:
\beq
\bI_1^{(0)}(\{p\}_m;\ep) = 
	\frac{\as}{2\pi}\frac{\Sep}{\Fep}
	\bigg(\frac{\mu^2}{Q^2}\bigg)^\ep
	\sum_{i=1}^{m} \bigg(
-\frac{1}{\ep^2} \sum_{\substack{k=1 \\ k\ne i}}^m \bT_i \ldot \bT_k
	+ \frac{1}{\ep} \gamma_{f_i}
	\bigg) y_{ik}^{-\ep} + \Oe{0}\,.
\label{eq:I10poles}
\eeq
It is straightforward to check that the poles of this expression coincide with those of 
the $\bI(\{p\};\ep)$ operator of \refr{Catani:1996vz}, hence $\int_1\dsiga{R}{1}{m+1}$ 
as given in \eqn{eq:INTsigmaRdiffA1} correctly cancels all $\ep$-poles of the virtual 
cross section $\dsig{V}{m}$. Thus the regularized virtual contribution,
\beq
\tsig{NLO}{m}[J] = 
	\int_{m} \Big[\dsig{V}{m} + \int_1 \dsiga{R}{1}{m+1}\Big]_{d=4} J_{m}\,,
\label{eq:sigmaNLO_m}
\eeq
is finite and integrable in four dimensions.

%%%
%%% The NNLO correction
%%%

\subsection{The NNLO correction}
\label{sec:colorfulNNLO-NNLO}

The NNLO correction to the cross section is a sum of three contributions, the 
tree level double real radiation, the one-loop plus a single radiation and the 
two-loop double virtual terms,
\beq
\tsig{NNLO}{}[J] = 
	\int_{m+2}\dsig{RR}{m+2} J_{m+2} 
	+ \int_{m+1}\dsig{RV}{m+1} J_{m+1} 
	+ \int_m\dsig{VV}{m} J_m\,,
\label{eq:sigmaNNLO} 
\eeq
which are all divergent in four dimensions. In the \colorfulNNLO\ method, we render 
these terms finite by the rearrangement
\beq 
\tsig{NNLO}{}[J] = 
	\int_{m+2}\dsig{NNLO}{m+2} 
	+ \int_{m+1}\dsig{NNLO}{m+1} 
	+ \int_m\dsig{NNLO}{m}\,,
\label{eq:sigmaNNLOfin} 
\eeq 
where,
\bal
\dsig{NNLO}{m+2} &= 
	\Big\{\dsig{RR}{m+2} J_{m+2} 
	- \dsiga{RR}{2}{m+2} J_{m} 
	-\Big[\dsiga{RR}{1}{m+2} J_{m+1} 
	- \dsiga{RR}{12}{m+2} J_{m}\Big]\Big\}_{d=4}\,, 
\label{eq:sigmaNNLOm+2} 
\\ 
\dsig{NNLO}{m+1} &= 
	\Big\{\Big[\dsig{RV}{m+1} 
	+ \int_1\dsiga{RR}{1}{m+2}\Big] J_{m+1}  
	-\Big[\dsiga{RV}{1}{m+1} 
	+ \Big(\int_1\dsiga{RR}{1}{m+2}\Big)\strut^{{\rm A}_{\scriptscriptstyle 1}} 
	\Big] J_{m} \Big\}_{d=4}\,,
\label{eq:sigmaNNLOm+1} 
\\
\dsig{NNLO}{m} &= 
	\Big\{\dsig{VV}{m} 
	+ \int_2\Big[\dsiga{RR}{2}{m+2} 
	- \dsiga{RR}{12}{m+2}\Big] 
	+\int_1\Big[\dsiga{RV}{1}{m+1} 
	+ \Big(\int_1\dsiga{RR}{1}{m+2}\Big) \strut^{{\rm A}_{\scriptscriptstyle 1}} 
	\Big]\Big\}_{d=4} J_{m}\,.
\label{eq:sigmaNNLOm}
\eal
The right-hand sides of \eqns{eq:sigmaNNLOm+2}{eq:sigmaNNLOm+1}
are integrable in four dimensions by construction
\cite{Somogyi:2005xz,Somogyi:2006da,Somogyi:2006db},
while the integrability in four dimensions of \eqn{eq:sigmaNNLOm}
is ensured by the KLN theorem.

Equation~\eqref{eq:sigmaNNLOm+2} includes the double real (RR) contribution that is singular 
whenever one or two partons become unresolved. In order to regularize the two-parton 
singularities, we subtract an approximate cross section,
\beq
\dsiga{RR}{2}{m+2} = \frac{1}{2s} \PS{m+2}(Q^2) \cA_2 \SME{m+2}{(0)}{}\,,
\eeq
where the double unresolved approximate matrix element for processes with $m+2$ partons 
in the final state is \cite{Somogyi:2006da}
\beq
\bsp
\cA_2 \SME{m+2}{(0)}{} &=
\sum_{r=1}^{m+2}\sum_{s=1}^{m+2}\Bigg\{
\sum_{\substack{i=1 \\ i\ne r,s}}^{m+2}\Bigg[\frac16\, \cC{irs}{(0,0)}
+ \sum_{\substack{j=1 \\ j\ne i,r,s}}^{m+2} \frac18\, \cC{ir;js}{(0,0)}
\\ 
&\quad +\, \frac12\,\Bigg( \cSCS{ir;s}{(0,0)}
- \cC{irs}{}\cSCS{ir;s}{(0,0)} 
- \sum_{\substack{j=1 \\ j\ne i,r,s}}^{m+2} \cC{ir;js}{} \cSCS{ir;s}{(0,0)} \Bigg)
\\ 
&\quad -\,\, \cSCS{ir;s}{}\cS{rs}{(0,0)}
- \frac12\, \cC{irs}{}\cS{rs}{(0,0)}
+ \cC{irs}{}\cSCS{ir;s}{}\cS{rs}{(0,0)}
\\ 
&\quad + \sum_{\substack{j=1 \\ j\ne i,r,s}}^{m+2} 
	\frac12\, \cC{ir;js}{}\cS{rs}{(0,0)}\Bigg] + \frac12\, \cS{rs}{(0,0)}
\Bigg\}
\,.
\label{eq:RR_A2}
\esp
\eeq
The functions $\cC{irs}{(0,0)}$, $\cC{ir;js}{(0,0)}$, $\cSCS{ir;s}{(0,0)}$ and 
$\cS{rs}{(0,0)}$ in \eqn{eq:RR_A2} are subtraction terms which regularize 
the ${\vec p}_i||{\vec p}_r||{\vec p}_s$ triple collinear, the ${\vec p}_i||{\vec p}_r$, 
${\vec p}_j||{\vec p}_s$ double collinear, the ${\vec p}_i||{\vec p}_r$, $p_s^\mu\to 0$ 
one collinear, one soft (collinear+soft) and the $p_r^\mu\to 0$, $p_s^\mu\to 0$ double 
soft limits. The rest of the counterterms appearing in \eqn{eq:RR_A2} account for 
the double or triple overlap of limits, hence multiple subtractions are avoided in 
overlapping double unresolved regions. The role of each specific counterterm is suggested 
by the notation. For instance, $\cC{irs}{}\cSCS{ir;s}{(0,0)}$ accounts for the triple 
collinear limit of the collinear+soft counterterm, with the rest of the counterterms 
having similar interpretations. All functions appearing in \eqn{eq:RR_A2} were defined 
explicitly in \refr{Somogyi:2006da}. The factors of $\frac16$, $\frac18$, etc., in 
\eqn{eq:RR_A2} appear so that each limit is counted precisely once, since the collinear 
indices of counterterms and the sums over them are not ordered in our convention.

After subtracting the double unresolved approximate cross section, the 
difference
\beq
\dsig{RR}{m+2} J_{m+2} - \dsiga{RR}{2}{m+2} J_{m}
\eeq
is still singular in the single unresolved regions of phase space. 
To regularize it, we also subtract
\beq
\dsiga{RR}{1}{m+2} = \frac{1}{2s} \PS{m+2}(Q^2) \cA_1 \SME{m+2}{(0)}{}\,,
\eeq
where $\cA_1$ has been defined in \eqn{eq:A1def}. To avoid double
subtraction in overlapping single and double unresolved regions of
phase space, we must also consider
\beq
\dsiga{RR}{12}{m+2} = \frac{1}{2s} \PS{m+2}(Q^2) \cA_{12} \SME{m+2}{(0)}{}\,,
\eeq
where the iterated single unresolved approximate matrix element reads
\beq
\cA_{12} \SME{m+2}{(0)}{} =
\sum_{t=1}^{m+2}\Bigg[
\sum_{\substack{k=1 \\ k\ne t}}^{m+2} \frac12 \,\cC{kt}{} \cA_2 \SME{m+2}{(0)}{}
+\Bigg(\cS{t}{} \cA_2 \SME{m+2}{(0)}{} -
\sum_{\substack{k=1 \\ k\ne t}}^{m+2} \cC{kt}{}\cS{t}{} \cA_2 \SME{m+2}{(0)}{} \Bigg) 
\Bigg]\,,
\label{eq:RR_A12}
\eeq
with the three terms above given by~\cite{Somogyi:2006da},
\bal
\cC{kt}{} \cA_2 & = 
\sum_{\substack{r=1 \\ r\ne k,t}}^{m+2} 
\Bigg[\cC{kt}{}\cC{ktr}{(0,0)} +\cC{kt}{}\cSCS{kt;r}{(0,0)} 
- \cC{kt}{}\cC{ktr}{}\cSCS{kt;r}{(0,0)}
- \cC{kt}{}\cC{rkt}{}\cS{kt}{(0,0)}
\nt\\ &\qquad\qquad
+ \sum_{\substack{i = 1 \\ i\ne r,k,t}}^{m+2}\Bigg(
\frac12\,\cC{kt}{}\cC{ir;kt}{(0,0)} 
- \cC{kt}{}\cC{ir;kt}{}\cSCS{kt;r}{(0,0)}\Bigg)\Bigg]
+ \cC{kt}{}\cS{kt}{(0,0)} \,,
\label{eq:CktA2}
\\
\cS{t}{} \cA_2 & =
\sum_{\substack{r=1 \\ r\ne t}}^{m+2}\Bigg\{\sum_{\substack{i = 1 \\ i\ne r,t}}^{m+2} 
\Bigg[\frac12\Bigg(\cS{t}{}\cC{irt}{(0,0)} +\cS{t}{}\cSCS{ir;t}{(0,0)} 
- \cS{t}{}\cC{irt}{}\cSCS{ir;t}{(0,0)}\Bigg)
\nt\\ &\qquad\qquad\qquad
- \cS{t}{}\cC{irt}{}\cS{rt}{(0,0)} - \cS{t}{}\cSCS{ir;t}{}\cS{rt}{(0,0)} 
+ \cS{t}{}\cC{irt}{}\cSCS{ir;t}{}\cS{rt}{(0,0)}\Bigg] 
+ \cS{t}{}\cS{rt}{(0,0)}\Bigg\} \,,
\label{eq:StA2}
\\
\cC{kt}{}\cS{t}{} \cA_2 & =
  \sum_{\substack{r=1 \\ r\ne k,t}}^{m+2} \Bigg[\cC{kt}{}\cS{t}{}\cC{krt}{(0,0)} 
+ \sum_{\substack{i = 1 \\ i\ne r,k,t}}^{m+2}\Bigg(
\frac12\cC{kt}{}\cS{t}{}\cSCS{ir;t}{(0,0)} 
- \cC{kt}{}\cS{t}{}\cSCS{ir;t}{}\cS{rt}{(0,0)}\Bigg)
\nt\\ &\qquad\qquad
- \cC{kt}{}\cS{t}{}\cC{krt}{}\cS{rt}{(0,0)}
- \cC{kt}{}\cS{t}{}\cC{rkt}{}\cS{kt}{(0,0)}
+ \cC{kt}{}\cS{t}{}\cS{rt}{(0,0)}\Bigg]
+ \cC{kt}{}\cS{t}{}\cS{kt}{(0,0)}\,.
\label{eq:CktStA2}
\eal
The notation in \eqnss{eq:CktA2}{eq:CktStA2} above serves to suggest the 
interpretation of the various terms. For instance, $\cC{kt}{}\cC{ktr}{(0,0)}$ 
in \eqn{eq:CktA2} accounts for the ${\vec p}_k||{\vec p}_t$ single collinear 
limit of the $\cC{ktr}{(0,0)}$ triple collinear counterterm, while, for example, 
$\cS{t}{}\cC{irt}{(0,0)}$ in \eqn{eq:StA2} represents the counterterm appropriate 
to the $p_t^\mu\to 0$ soft limit of $\cC{irt}{(0,0)}$. 
Clearly, $\cA_{12} \SME{m+2}{(0)}{}$ cancels the single unresolved singularities 
of the double unresolved subtraction term $\cA_{2} \SME{m+2}{(0)}{}$ by construction. 
Moreover, very importantly, $\cA_{12} \SME{m+2}{(0)}{}$ cancels at the same time the 
double unresolved singularities of the single unresolved subtraction term 
$\cA_{1} \SME{m+2}{(0)}{}$, as shown in \cite{Somogyi:2006da}. Hence the overlap of 
single and double unresolved subtractions is properly taken into account.
All of the counterterms appearing in \eqnss{eq:CktA2}{eq:CktStA2} were defined in 
\refr{Somogyi:2006da} explicitly. As before, the collinear indices and sums over 
them in \eqnss{eq:RR_A12}{eq:CktStA2} are not ordered,  so factors of $\frac12$ appear 
at various instances. The combination of terms appearing in \eqn{eq:sigmaNNLOm+2} was 
shown to be integrable in all kinematic limits in \refr{Somogyi:2006da}. Thus, the 
regularized double real contribution to the $m$-jet cross section
is finite and can be computed numerically in four dimensions for any infrared-safe 
observable.

Turning to \eqn{eq:sigmaNNLOm+1}, it describes the emission at one loop of one additional 
parton, the real-virtual (RV) contribution. In addition to explicit $\ep$-poles coming 
from the one-loop matrix element, the RV contribution has kinematical singularities when 
the additional parton becomes unresolved. The explicit poles are cancelled by the integral 
of the single unresolved subtraction term in the double real emission contribution to the 
full NNLO cross section, which is simply given by \eqns{eq:INTsigmaRdiffA1}{eq:I10} after 
the obvious replacement of $m\to m+1$
\beq
\int_1\dsiga{RR}{1}{m+2} = 
	\dsig{R}{m+1} \otimes \bI_1^{(0)}(\{p\}_{m+1};\ep)\,.
\label{eq:INTsigmaRRdiffA1}
\eeq
As shown above in \eqn{eq:I10poles} the combination,
\beq
\dsig{RV}{m+1} + \int_1 \dsiga{RR}{1}{m+1}
\label{eq:RVm+1plusI1RR1m+2}
\eeq
is finite in $\ep$. 
Nevertheless, \eqn{eq:RVm+1plusI1RR1m+2} is still singular in the single unresolved 
regions of phase space and requires regularization. We achieve this by subtracting 
two suitably defined approximate cross sections, 
$\dsiga{RV}{1}{m+1}$ and $\left(\int_1\dsiga{RR}{1}{m+2}\right)^{\rm{A}_1}$.
First, we consider
\beq
\dsiga{RV}{1}{m+1} = 
	\frac{1}{2s} \PS{m+1}(Q^2)
	 \cA_1 
	 2\Re\braket{m+1}{(0)}{}{(1)}
	 \,,
\label{eq:sigmaRVA1diff}
\eeq
which matches the kinematic singularity structure of $\dsig{RV}{m+1}$.
The general definition of the real-virtual counterterm is~\cite{Somogyi:2006db}
\beq
\bsp
\cA_1 
	 2\Re\braket{m+1}{(0)}{}{(1)} &=
\sum_{r=1}^{m+1} \Bigg[
\sum_{\substack{i=1 \\ i\ne r}}^{m+1} \frac{1}{2} \cC{ir}{(0,1)}
+ \Bigg(\cS{r}{(0,1)} 
- \sum_{\substack{i=1 \\ i\ne r}}^{m+1} \cCS{ir}{r}{(0,1)}\Bigg) \Bigg]
\\ &
+ \sum_{r=1}^{m+1} \Bigg[
\sum_{\substack{i=1 \\ i\ne r}}^{m+1} \frac{1}{2} \cC{ir}{(1,0)}
+ \Bigg(\cS{r}{(1,0)} 
- \sum_{\substack{i=1 \\ i\ne r}}^{m+1} \cCS{ir}{r}{(1,0)}\Bigg) 
\Bigg]\,.
\label{eq:A11M0M1}
\esp
\eeq
The basic organization of this subtraction in terms of unresolved limits is 
identical to the tree level single unresolved counterterm in \eqn{eq:A1def}. However 
in \eqn{eq:A11M0M1} we have terms with tree level collinear or soft functions multiplying 
(in color or spin space) one-loop matrix elements (those with the $(0,1)$ superscript),
as well as terms with one-loop collinear or soft functions multiplying tree level 
matrix elements (denoted with the $(1,0)$ superscript). This reflects the structure of 
infrared factorization of one-loop QCD matrix elements 
\cite{Bern:1997sc,Kosower:1999xi,Kosower:1999rx,Bern:1999ry}. The functions appearing 
in \eqn{eq:A11M0M1} are defined explicitly in \refr{Somogyi:2006db}. 

Then we consider the counterterm,
\beq
\Big(\int_1\dsiga{RR}{1}{m+2}\Big)\strut^{\rm{A}_1} =
	\frac{1}{2s} \PS{m+1}(Q^2) \cA_1 
	\left(\SME{m+1}{(0)}{} \otimes \bI_1^{(0)} \right)\,,
\label{eq:sigmaRRdiffA1A1}
\eeq
which regularizes the kinematic singularities of $\int_1\dsiga{RR}{1}{m+2}$.
This counterterm is given by~\cite{Somogyi:2006db}
\beq
\bsp
 \cA_1  \Big(\SME{m+1}{(0)}{} \otimes \bI_1^{(0)} \Big) &=
\sum_{r=1}^{m+1} \Bigg[
\sum_{\substack{i=1 \\ i\ne r}}^{m+1} \frac{1}{2} \cC{ir}{(0,0 \otimes I)}
+ \Bigg(\cS{r}{(0,0 \otimes I)}
- \sum_{\substack{i=1 \\ i\ne r}}^{m+1} \cCS{ir}{r}{(0,0 \otimes I)}\Bigg) 
\Bigg]
\\&
+ \sum_{r=1}^{m+1} \Bigg[
\sum_{\substack{i=1 \\ i\ne r}}^{m+1} \frac{1}{2} \cC{ir}{R\times(0,0)}
+ \Bigg(\cS{r}{R\times(0,0)}
- \sum_{\substack{i=1 \\ i\ne r}}^{m+1} \cCS{ir}{r}{R\times(0,0)}\Bigg) 
\Bigg]\,.
\label{eq:A10M0I}
\esp
\eeq
The structure of this subtraction in terms of unresolved limits is again the
same as the tree level single unresolved counterterm in \eqn{eq:A1def}. 
However we have two types of terms for each limit, labeled by the different 
superscripts. The reason is the following. This counterterm is constructed from 
factorization formul\ae\ describing the behavior of the product 
of a QCD squared matrix element times the $\bI_1^{(0)}$ insertion operator of 
\eqn{eq:I10} in the collinear and soft limits. (The existence of a 
universal collinear factorization formula for the product $\SME{m+1}{(0)}{} \otimes 
\bI_1^{(0)}$ is not guaranteed by the factorization properties of QCD matrix elements. 
The requirement that such a formula exists puts highly non-trivial constraints on the 
form of $\bI_1^{(0)}$, i.e., on the specific definition of the single unresolved approximate 
cross section. See section~4.1.1 of \refr{Somogyi:2006db} for a discussion of this point.) 
These factorization formul\ae\ were computed in \refr{Somogyi:2006db} and turn out to be 
sums of two pieces. Both pieces involve the product of a tree level collinear or soft 
function times a tree level matrix element. One piece is further multiplied by the 
$\bI_1^{(0)}$ insertion operator appropriate to the reduced matrix element, while the other 
is multiplied with a well-defined scalar (in color space) remainder function $R$. The 
superscripts on the various terms in \eqn{eq:A10M0I} are meant to reflect this structure. 
The combination of terms appearing in \eqn{eq:sigmaNNLOm+1} is both free of $\ep$-poles 
and integrable in all kinematically singular limits \cite{Somogyi:2006db}. Hence, the 
regularized real-virtual contribution to the $m$-jet cross section
is finite and can be computed numerically in four dimensions for any infrared-safe 
observable.

Finally, the two-loop double virtual (VV) contribution to the NNLO corrections appears 
in \eqn{eq:sigmaNNLOm}. The VV contribution has explicit infrared poles that
cancel against the poles of the four integrated counterterms, which are shown in 
\eqn{eq:sigmaNNLOm}. The integral of the real-virtual counterterms (the 
last two terms of \eqn{eq:sigmaNNLOm}) was computed in 
\refrs{Somogyi:2008fc,Aglietti:2008fe,Bolzoni:2009ye} and can be written as
\beq
\int_{1}\dsiga{RV}{1}{m+1} =
  \dsig{V}{m}\otimes \bI_{1}^{(0)}(\mom{}_{m};\ep)
+ \dsig{B}{m}\otimes \bI_{1}^{(1)}(\mom{}_{m};\ep)
\label{eq:I1dsigRVA1}
\eeq
and 
\beq
\int_1\Big(\int_1\dsiga{RR}{1}{m+2}\Big)
\strut^{{\rm A}_{\scriptscriptstyle 1}} =
\dsig{B}{m}\otimes \left[
\frac12\Big\{
\bI_{1}^{(0)}(\mom{}_{m};\ep),
\bI_{1}^{(0)}(\mom{}_{m};\ep)
\Big\}
+ \bI_{1,1}^{(0,0)}(\mom{}_{m};\ep)
\right]
\,.
\label{eq:I1dsigRA1_A1}
\eeq
The insertion operator $\bI_{1}^{(0)}$ is given in \eqn{eq:I10},
while $\bI_{1}^{(1)}$ and $\bI_{1,1}^{(0,0)}$ have the following color
decompositions:
\beq
\bsp
\bI_{1}^{(1)}(\mom{}_{m};\ep) &=
     \left[\frac{\as}{2\pi}\frac{\Sep}{\Fep} \left(\frac{\mu^2}{Q^2}\right)^\ep\right]^2
     \sum_{i}\bigg[
     \IcC{1,\fla{i}}{(1)}(y_{iQ};\ep)\,\CA \bTsq{i}
     +\sum_{k\ne i} \IcS{1}{(1),(i,k)}(\Y{i}{k};\ep)\,\CA \bT_{i}\bT_{k}
\\ &
     +\sum_{k\ne i} \sum_{l\ne i,k}
     \IcS{1}{(1),(i,k,l)}(\Y{i}{k},\Y{i}{l},\Y{k}{l};\ep)
     \sum_{a,b,c}f_{abc}\bT_i^a \bT_k^b \bT_l^c
     \bigg]
\esp
\label{eq:I11}
\eeq
and
\beq
\bI_{1,1}^{(0,0)}(\mom{}_{m};\ep) =
     \left[\frac{\as}{2\pi}\frac{\Sep}{\Fep} \left(\frac{\mu^2}{Q^2}\right)^\ep\right]^2
     \sum_{i}\bigg[
     \IcC{1,1,\fla{i}}{(0,0)}(y_{iQ};\ep)\,\CA\bTsq{\fl{i}}
     +\sum_{k\ne i}
     \IcS{1,1}{(0,0),(i,k)}(\Y{i}{k};\ep)\,\CA\,\bT_{i}\bT_{k}
\bigg]\,.
\label{eq:I1100}
\eeq
Again, there is no one-to-one correspondence between the unintegrated double
unresolved subtraction terms in \eqns{eq:A11M0M1}{eq:A10M0I} and the
kinematic functions that appear in \eqns{eq:I11}{eq:I1100}. The latter
are obtained from the former after integration over unresolved momenta
and summation over unobserved colors and flavors. This remark applies
also to the rest of the insertion operators to be discussed below.

The integral of the iterated single unresolved counterterm (the third term
of \eqn{eq:sigmaNNLOm}) was evaluated in \refr{Bolzoni:2010bt} yielding the result
\beq
\int_2 \dsiga{RR}{12}{m+2} = 
	\dsig{B}{m} \otimes \bI_{12}^{(0)}(\{p\}_m;\ep)\,.
\label{eq:INTsigmaRRA12diff}
\eeq
The insertion operator has five contributions according to 
the possible color structures,
\beq 
\bsp 
\bI_{12}^{(0)}(\{p\};\ep) &=  
	\left[\frac{\as}{2\pi} \frac{\Sep}{\Fep}
     \left(\frac{\mu^2}{Q^2}\right)^{\ep}\right]^2 
	\Bigg\{ \sum_{i=1}^m \Bigg[ 
	\IcC{12,i}{(0)}(y_{iQ};\ep) \, \bT_i^2 
	+ \sum_{\substack{j=1 \\ j\ne i}}^m 
	\IcC{12,i j}{(0)}(y_{iQ},y_{jQ},Y_{ij,Q};\ep)\, \bT_j^2 \Bigg] 
	\bT_i^2 
\\ & \quad 
	+ \sum_{\substack{j,l=1 \\ l\ne j}}^m \Bigg[ 
	\IcS{12}{(0),(j,l)}(Y_{jl,Q};\ep)\, \CA\,  
	+ \sum_{i=1}^m \IcSCS{12,i}{(0),(j,l)}(y_{iQ},Y_{ij,Q},Y_{il,Q},Y_{jl,Q};\ep)\, 
	\bT_i^2\, \Bigg] \bT_{j} \ldot \bT_{l} 
\\ & \quad
	+ \sum_{\substack{i,k=1, \\ k\ne i}}^m\sum_{\substack{j,l=1, \\ l\ne j}}^m 
	\IcS{12}{(0),(i,k)(j,l)}
	(Y_{ik,Q},Y_{ij,Q},Y_{il,Q},Y_{jk,Q},Y_{kl,Q},Y_{jl,Q};\ep) 
	\{ \bT_{i} \ldot \bT_{k},\bT_{j} \ldot \bT_{l} \} 
	\Bigg\} \,.
\label{eq:I12} 
\esp 
\eeq 

Finally, the integration of the collinear-type contributions to the double 
unresolved counterterm (the second term of \eqn{eq:sigmaNNLOm}) was performed 
in \refr{DelDuca:2013kw}.  The soft-type contributions to the same integral 
were presented in \refr{Somogyi:2013yk}. We find
\beq
\int_2 \dsiga{RR}{2}{m+2} = 
	\dsig{B}{m} \otimes \bI_{2}^{(0)}(\{p\}_m;\ep)\,,
\label{eq:INTsigmaRRA2diff}
\eeq
where the structure of the insertion operator $\bI_{2}^{(0)}$ is identical 
to $\bI_{12}^{(0)}$ in color space,
\beq 
\bsp 
\bI_{2}^{(0)}(\{p\};\ep) &=  
	\left[\frac{\as}{2\pi} \frac{\Sep}{\Fep}
     \left(\frac{\mu^2}{Q^2}\right)^{\ep}\right]^2 
	\Bigg\{ \sum_{i=1}^m \Bigg[ 
	\IcC{2,i}{(0)}(y_{iQ};\ep) \, \bT_i^2 
	+ \sum_{\substack{j=1 \\ j\ne i}}^m 
	\IcC{2,i j}{(0)}(y_{iQ},y_{jQ},Y_{ij,Q};\ep)\, \bT_j^2 \Bigg] 
	\bT_i^2 
\\ & \quad 
	+ \sum_{\substack{j,l=1 \\ l\ne j}}^m \Bigg[ 
	\IcS{2}{(0),(j,l)}(Y_{jl,Q};\ep)\, \CA\,  
	+ \sum_{i=1}^m \IcSCS{2,i}{(0),(j,l)}(y_{iQ},Y_{ij,Q},Y_{il,Q},Y_{jl,Q};\ep)\, 
	\bT_i^2\, \Bigg] \bT_{j} \ldot \bT_{l} 
\\ & \quad
	+ \sum_{\substack{i,k=1, \\ k\ne i}}^m\sum_{\substack{j,l=1, \\ l\ne j}}^m 
	\IcS{2}{(0),(i,k)(j,l)}
	(Y_{ik,Q},Y_{ij,Q},Y_{il,Q},Y_{jk,Q},Y_{kl,Q},Y_{jl,Q};\ep) 
	\{ \bT_{i} \ldot \bT_{k},\bT_{j} \ldot \bT_{l} \} 
	\Bigg\} \,.
\label{eq:I2} 
\esp 
\eeq 

The kinematic functions entering the various insertion operators in eqs.\ (\ref{eq:I10}), 
(\ref{eq:I11}), (\ref{eq:I1100}), (\ref{eq:I12}) and (\ref{eq:I2}) have been expanded 
in $\ep$. The coefficients of the poles in these Laurent expansions have been computed fully 
analytically. The resulting expressions are rather lengthy and involve, in addition 
to logarithms, dilogarithms and trilogarithms of rational arguments in the variables 
$y_{iQ}$ and $Y_{jk,Q}$. For the finite parts, we computed analytically all terms that diverge 
logarithmically on the boundaries of the phase space (i.e., when $y_{iQ}\to 0$ and/or 
$Y_{jk,Q}\to 0$), while the remaining regular contributions were computed numerically. 
We stress that our method is generic and we can construct counterterms 
for processes with an arbitrary number $m$ of jets in the final state.
The only missing ingredients are the corresponding two-loop matrix elements, 
and current only the two-loop matrix elements for two and three-jet production are available.
Since the poles of all integrated counterterms are known analytically, we can 
demonstrate explicitly that the regularized double virtual contribution to the $m$-jet 
cross section is finite and free of $\ep$-poles. For $m=2$ this was done in 
\refr{DelDuca:2015zqa}, while the $m=3$ case will be discussed in the next section.

%%%
%%% Electron-positron annihilation into three jets}
%%%
% ========== ========== ========== ========== ==========

\section{Electron-positron annihilation into three jets}
\label{sec:ee3jets}

We consider $e^+e^-\to$~3~jet production through the  exchange of a photon or a $Z$ boson
of momentum $Q$ in the $s$ channel.
Through NNLO in QCD, this production cross section receives contributions from
the following partonic subprocesses:
\\
\begin{center}\begin{tabular}{cll}
LO & $\gamma^*/Z^*(Q) \to q(p_1) + \qb(p_2) + g(p_3)$ & tree level
\\[0.5em]
NLO & $\gamma^*/Z^*(Q) \to q(p_1) + \qb(p_2) + g(p_3) + g(p_4)$ & tree level
\\ & $\gamma^*/Z^*(Q) \to q(p_1) + \qb(p_2) + q'(p_3) + \qb'(p_4)$ & tree level
\\ & $\gamma^*/Z^*(Q) \to q(p_1) + \qb(p_2) + q(p_3) + \qb(p_4)$ & tree level
\\ & $\gamma^*/Z^*(Q) \to q(p_1) + \qb(p_2) + g(p_3)$ & one-loop
\\[0.5em]
NNLO & $\gamma^*/Z^*(Q) \to q(p_1) + \qb(p_2) + g(p_3) + g(p_4) + g(p_5)$ & tree level
\\ & $\gamma^*/Z^*(Q) \to q(p_1) + \qb(p_2) + q'(p_3) + \qb'(p_4) + g(p_5)$ & tree level
\\ & $\gamma^*/Z^*(Q) \to q(p_1) + \qb(p_2) + q(p_3) + \qb(p_4) + g(p_5)$ & tree level
\\ & $\gamma^*/Z^*(Q) \to q(p_1) + \qb(p_2) + g(p_3) + g(p_4)$ & one-loop
\\ & $\gamma^*/Z^*(Q) \to q(p_1) + \qb(p_2) + q'(p_3) + \qb'(p_4)$ & one-loop
\\ & $\gamma^*/Z^*(Q) \to q(p_1) + \qb(p_2) + q(p_3) + \qb(p_4)$ & one-loop
\\ & $\gamma^*/Z^*(Q) \to q(p_1) + \qb(p_2) + g(p_3)$ & two-loop
\end{tabular}
\end{center}
~\\
where we show the four-momenta of the particles in parentheses.
The tree level matrix elements for the production of five jets were first 
obtained in \refrs{Hagiwara:1988pp,Berends:1988yn,Falck:1989uz}, while the 
one-loop corrections to four jet production have been computed in 
\refrs{Glover:1996eh,Bern:1996ka,Campbell:1997tv,Bern:1997sc}. The two-loop matrix 
elements for $\gamma^*/Z^* \to q \qb g$ are also available both in squared matrix 
element form \cite{Garland:2001tf} and as helicity amplitudes \cite{Garland:2002ak}. 
In the \colorfulNNLO\ framework the subtraction terms correctly account for all spin 
and color correlations in the various infrared limits. Hence, we also need the 
three-parton and four-parton matrix elements including color and/or spin correlations. 
When there are only three partons in the final state, the color correlations factorize 
completely (see \eqn{eq:3colors} below), so computing the color-correlated three-parton 
matrix elements is trivial at any loop order. This is no longer the case for the 
four-parton matrix elements. In our computation, we only need the four-parton 
color-correlated matrix elements at tree-level and these are given in 
\refr{Nagy:1998bb}\footnote{Note a misprint in eqs.\ (B.11)--(B.13) of \refr{Nagy:1998bb}: 
the 2, 3 and 4 indices of the $M_0^{ik}$, $M_x^{ik}$ and $M_{xx}^{ik}$ matrices should be 
cyclicly permuted, $(2,3,4) \to (4,2,3)$.}. 
The required spin-correlated matrix elements on the other hand are rather easy to 
implement starting from helicity amplitudes. 

The sum of the three-, four- and five-parton contributions is finite for any infrared-safe 
observable, but the four- and five-parton contributions to three-jet observables contain infrared 
singularities associated with unresolved real emission, which must be subtracted and 
cancelled against the infrared singularities coming from loop integrals in the three- and 
four-parton final states. We accomplish this cancellation with the \colorfulNNLO\ 
method as outlined in the previous section. 

%%%
%%% $e^+e^-\to$~3~jet production at NLO
%%%

\subsection{$e^+e^-\to$~3~jet production at NLO}
\label{sec:ee3jetsnlo}

It is instructive to spell out the computation of the NLO correction in some detail. 
The four-parton real emission contribution to the differential cross section for 
three-jet production is
\beq
\dsig{R}{4} = 
	\frac{1}{2s} \PS{4}(Q^2)\,
	 \sum_q \Bigg( 
	 \frac{1}{2!} \SME{q\qb gg}{(0)}{} 
	 + \sum_{q' \ne q} \SME{q\qb q'\qb'}{(0)}{} 
	 + \frac{1}{(2!)^2} \SME{q\qb q\qb}{(0)}{} \Bigg) \,.
\label{eq:sigma3jRdiff}
\eeq
The integral over the phase space is divergent in four dimensions because of the 
singularities in the regions where one parton is collinear and/or soft. In order
to regularize those singularities, we subtract
\beq
\dsiga{R}{1}{4} = 
	\frac{1}{2s} \PS{4}(Q^2)\,
	 \sum_q \Bigg( 
	 \frac{1}{2!}  \cA_1 \SME{q\qb gg}{(0)}{} 
	 + \sum_{q' \ne q}  \cA_1 \SME{q\qb q'\qb'}{(0)}{} 
	 + \frac{1}{(2!)^2}  \cA_1 \SME{q\qb q\qb}{(0)}{} \Bigg) \,,
\label{eq:sigma3jRA1diff}
\eeq
where the approximate matrix elements are defined in \eqn{eq:A1def}.
The counterterms are explicitly defined in \refrs{Somogyi:2006cz,Somogyi:2006da} 
in a form that is immediately suitable for inclusion in a general purpose 
computer code. By generating sequences of phase space points tending to each 
infrared limit, we have checked that the sum of subtractions correctly reproduces 
the real emission differential cross section point-by-point in any single unresolved 
region of phase space. As a consequence the difference
\beq
\dsig{NLO}{4}\equiv \dsig{R}{4} J_4 - \dsiga{R}{1}{4} J_3\,,
\label{eq:dsigNLO_4}
\eeq
is integrable in four dimensions and the regularized real contribution
can be computed using whatever numerical procedure is most convenient.

Turning to the three-parton virtual contribution, we have
\beq
\dsig{V}{3} = 
	\frac{1}{2s} \PS{3}(Q^2)\, 
	\sum_q 
	2\Re\braket{q\qb g}{(0)}{}{(1)}
\,.
\label{eq:sigma3jVdiff}
\eeq
Equation~\eqref{eq:sigma3jVdiff} contains explicit $\ep$-poles coming from the one-loop matrix 
element. These poles are cancelled by adding back the approximate cross section that 
we have subtracted from the real correction in integrated form which can be written as 
in \eqn{eq:INTsigmaRdiffA1} (with $m=3$)
\beq
\int_1\dsiga{R}{1}{4} = 
	\dsig{B}{3} \otimes \bI_1^{(0)}(\{p\}_3;\ep)\,.
\eeq
The insertion operator $\bI_1^{(0)}$ is given in \eqn{eq:I10}. For three-jet production, 
as there are only three partons in the final state, the color connections that appear 
in the generic case in \eqn{eq:I10} factorize completely,
\beq
\bT_1\cdot \bT_2 = \frac{\CA}2 - \CF
\quad\mbox{and}\quad
\bT_1\cdot \bT_3 = \bT_2\cdot\bT_3 = -\frac{\CA}{2}\,.
\label{eq:3colors}
\eeq
Thus,
\beq
\bsp
\bI_1^{(0)}(\{p\}_3;\ep) &= 
	\frac{\as}{2\pi}\frac{\Sep}{\Fep}
	\bigg(\frac{\mu^2}{Q^2}\bigg)^\ep
	\bigg\{
	\CF\bigg[
		\IcC{1,q}{(0)}(y_{1Q};\ep)
		+ \IcC{1,q}{(0)}(y_{2Q};\ep)
		- 2\IcS{1}{(0),(1,2)}(Y_{12,Q};\ep) 
		\bigg]
\\&
	+\CA\bigg[
		\IcC{1,g}{(0)}(y_{3Q};\ep)
 		+ \IcS{1}{(0),(1,2)}(Y_{12,Q};\ep) 
		- \IcS{1}{(0),(1,3)}(Y_{13,Q};\ep) 
		- \IcS{1}{(0),(2,3)}(Y_{23,Q};\ep)
		\bigg]
	\bigg\}\,.
\esp
\label{eq:I103j}
\eeq
Using \eqn{eq:I10poles} (or the expressions in \appx{appx:I10}), it is straightforward to check that
\beq
\bsp
\bI_1^{(0)}(\{p\}_3;\ep) = 
	\frac{\as}{2\pi}\frac{\Sep}{\Fep}
	\bigg(\frac{\mu^2}{Q^2}\bigg)^\ep
	\bigg\{&
	\frac{2\CF + \CA}{\ep^2}	
	+ \frac{1}{\ep}
		\bigg[
		(\CA-2\CF)\Log{y_{12}} 
		- \CA (\Log{y_{13}} + \Log{y_{23}})
\\&
		+\frac{11}{6} \CA + 3\CF - \frac{2}{3}\Nf \TR
		\bigg]
	+ \Oe{0}
	\bigg\}
\esp
\label{eq:I103jexp}
\eeq
and that the combination
\beq
\dsig{NLO}{3} \equiv 
	\bigg[\dsig{V}{3} + \int_1 \dsiga{R}{1}{4}\bigg] J_3
\label{eq:dsigNLO_3jet}
\eeq
is free of $\ep$-poles. Thus \eqn{eq:dsigNLO_3jet} is finite in four dimensions 
and the regularized virtual contribution can be computed using standard numerical 
techniques for any infrared-safe observable.

%%%
%%% $e^+e^-\to$~3~jet production at NNLO
%%%

\subsection{$e^+e^-\to$~3~jet production at NNLO}
\label{sec:ee3jetsnnlo}

Turning to the NNLO correction, we consider first the double real emission contribution 
to the differential cross section for three-jet production,
\beq
\dsig{RR}{5} = 
	\frac{1}{2s} \PS{5}(Q^2)\,
	 \sum_q \Bigg( 
	 \frac{1}{3!} \SME{q\qb ggg}{(0)}{} 
	 + \sum_{q' \ne q} \SME{q\qb q'\qb' g}{(0)}{} 
	 + \frac{1}{(2!)^2} \SME{q\qb q\qb g}{(0)}{} \Bigg) \,.
\label{eq:sigma3jRRdiff}
\eeq
The integral over the phase space is divergent in four dimensions because of infrared 
singularities in regions of phase space where one or two partons are collinear and/or soft. 
In order to regularize those singularities, we subtract approximate cross sections
$\dsiga{RR}{2}{5}$, $\dsiga{RR}{1}{5}$ and $\dsiga{RR}{12}{5}$ as explained in 
\sect{sec:colorfulNNLO-NNLO}. The counterterms are defined in \refr{Somogyi:2006da} 
explicitly, in a form directly suited for implementation into a general purpose computer 
code. We have checked in all kinematic limits that the difference
\beq
\dsig{NNLO}{5} \equiv
	\dsig{RR}{5} J_5 - \dsiga{RR}{2}{5} J_3
	- \dsiga{RR}{1}{5} J_4 + \dsiga{RR}{12}{5} J_3
\label{eq:dsigNNLO_5}
\eeq
is integrable in four dimensions by generating sequences of phase space points tending 
to each infrared limit. Hence the double real emission differential cross section is 
regularized point-by-point in phase space. The complete locality of the subtractions 
then ensures that the integral of \eqn{eq:dsigNNLO_5} is well-defined and finite in four 
dimensions for any infrared-safe observable and can be computed with any suitable numerical 
technique.

The real-virtual contribution to the differential cross section is
\beq
\bsp
\dsig{RV}{4} = 
	\frac{1}{2s} \PS{4}(Q^2)
	 \sum_q \bigg(&
	 \frac{1}{2!} 
	 	2\Re\braket{q\qb gg}{(0)}{}{(1)}
	 + \sum_{q' \ne q}  
	 	2\Re\braket{q\qb q'\qb'}{(0)}{}{(1)}
\\&
	 + \frac{1}{(2!)^2}  
	 	2\Re\braket{q\qb q\qb}{(0)}{}{(1)}
	\bigg) \,.
\label{eq:sigma3jRVdiff}
\esp
\eeq
Equation~\eqref{eq:sigma3jRVdiff} contains explicit $\ep$-poles coming from the one-loop 
matrix element and furthermore it is divergent in phase space regions where a parton becomes 
unresolved. The explicit poles are cancelled by the integral of the single unresolved 
subtraction term in the double real emission contribution to the full NNLO cross section, 
$\int_1 \dsiga{RR}{1}{5}$. The calculation in \refr{Somogyi:2006cz} for general $m$ assures 
us that the combination
\beq
\dsig{RV}{4} + \int_1 \dsiga{RR}{1}{5}
\label{eq:RV4plusI1RR15}
\eeq
is finite in $\ep$. (Of course, this can be checked explicitly using \eqn{eq:I10poles} 
or the expressions in \appx{appx:I10} as well.) However, \eqn{eq:RV4plusI1RR15} is still 
singular in the single unresolved regions of phase space. We regularize these singularities 
by subtracting the approximate cross sections $\dsiga{RV}{1}{4}$ and 
$\Big(\int_1\dsiga{RR}{1}{5}\Big)\strut^{\rm{A}_1}$ as discussed in 
\sect{sec:colorfulNNLO-NNLO}.
The explicit definitions of the counterterms in \refr{Somogyi:2006db} can be straightforwardly 
implemented into a computer code in a general way. It is then easy to check numerically that
the combination
\beq
\dsig{NNLO}{4}\equiv
	\bigg[\dsig{RV}{4} + \int_1 \dsiga{RR}{1}{5}\bigg] J_4
	-\bigg[\dsiga{RV}{1}{4} +\Big(\int_1\dsiga{RR}{1}{5}\Big)\strut^{\rm{A}_1}\bigg] J_3
\label{eq:dsigNNLO_4}
\eeq
is both free of $\ep$-poles and integrable in all kinematically singular
limits in four dimensions (as usual, by generating sequences of phase space points 
tending to all infrared limits). Thus, since the subtractions are fully local, the 
regularized real-virtual contribution to the 3-jet differential cross section 
is well-defined and finite and can be computed numerically in four dimensions for any 
infrared-safe observable.

Finally, the double virtual contribution to the differential cross section reads
\beq
\dsig{VV}{3} = 
\frac{1}{2s} \PS{3}(Q^2)\, 
\sum_q\left( 
	\SME{q\qb g}{(1)}{} 
	+
	2\Re\braket{q\qb g}{(0)}{}{(2)} 
	\right)\,,
\label{eq:sigmaVVdiff}
\eeq
and contains explicit $\ep$-poles coming from the two-loop matrix element
and the square of the one-loop matrix element. The structure of these poles 
was presented explicitly in \refr{Garland:2001tf} which we reproduce here using 
our conventions for the notation:
\beq
\bsp
\SME{q\qb g}{(1)}{} + 2\Re\braket{q\qb g}{(0)}{}{(2)} 
&=
	2\Re\braket{q\qb g}{(0)}{}{(1)} 
	\otimes 2\,\Re\bI_{q\qb g}^{(1)}(\ep)
- \SME{q\qb g}{(0)}{} \otimes
	2\Big(\Re \bI_{q\qb g}^{(1)}(\ep)\Big)^2
\\&
+ \SME{q\qb g}{(0)}{} \otimes \bigg[
  \e^{-\ep \gamma_E} \frac{\Gamma(1-2\ep)}{\ep \Gamma(1-\ep)}
  \Big(\beta_0 + 2\ep K\Big)\,\Re \bI_{q\qb g}^{(1)}(2\ep)
\\&\qquad\qquad\qquad
- \frac{\beta_0}{\ep}\,\Re \bI_{q\qb g}^{(1)}(\ep)
+  \frac{\Sep}{\Fep}\frac{1}{2 \ep} \Big(2 H_q(\Nf) + H_g(\Nf)\Big)
\bigg]
\\&
+ \Oe{0}\,,
\esp
\label{eq:VV3jpoles}
\eeq
where the universal constants are
\beq
K = \left(\frac{67}{18} - \frac{\pi^2}{6} \right)\CA - \frac{10}{9} \TR \Nf
\,,
\eeq 
\bal
H_q(\Nf) &= 
  \CA \CF \left(\frac{13 \zeta_3}{2}+\frac{245}{216}-\frac{23 \pi^2}{48}\right)
+ \CF^2 \left(-6 \zeta_3-\frac{3}{8}+\frac{\pi^2}{2}\right)\nonumber\\
&+ \CF \Nf \TR \left(\frac{\pi ^2}{12}-\frac{25}{54}\right)
\,,
\label{eq:Hq}
\\ 
H_g(\Nf) &= 
  \CA^2 \left(\frac{\zeta_3}{2}+\frac{5}{12}+\frac{11 \pi^2}{144}\right)
- \CA \Nf \TR\left(\frac{205}{54}+\frac{\pi ^2}{36}\right)
- \CF \Nf \TR+\frac{20}{27} \Nf^2 \TR^2\,,
\label{eq:Hg}
\eal
and the three-parton insertion operator is
\beq
\bI_{q\qb g}^{(1)}(s_{12}, s_{13}, s_{23},\mu^2;\ep) = 
\frac{\as}{4\pi}\frac{\Sep}{\Fep}
\bigg[
  \frac{1}{\ep^2}\sum_{i=1}^3 \sum_{\substack{k=1 \\ k\ne i}}^3
  \bigg(\frac{\mu^2}{-s_{ik}}\bigg)^{\ep}\,\bT_i \ldot \bT_k
- \frac{1}{\ep} (2\gamma_q + \gamma_g)
\bigg]\,,
\label{eq:I(1)}
\eeq
with
\beq
\gamma_q = \frac32\CF
\quad\mbox{and}\quad
\gamma_g = \frac{\beta_0}{2}
\,.
\eeq
The signs of the imaginary parts of the $(-s_{ik})^{-\ep}$ factors are
fixed by the usual $s_{ik}+\ri \varepsilon$ prescription on the
Feynman-propagators,
\beq 
\left(\frac{\mu^{2}}{-s_{ik}}\right)^{\ep} = 
\left(\frac{\mu^{2}}{\left|s_{ik}\right|}\right)^{\ep}
\left[
1+\left(\ri\pi\ep 
-\frac{\pi^{2}}{2}\ep^2\right)\Theta\left(s_{ik}\right) 
+\Oe{3}\right]\,.
\eeq
Hermitian conjugation flips the sign of the imaginary parts. We note that 
the poles of this operator are closely related to those of the 
$\bI_1^{(0)}(\{p\}_3;\ep)$ operator of \eqn{eq:I103jexp}:
\beq
\bI_1^{(0)}(\{p\}_3;\ep) = 
	-2\Re \bI_{q\qb g}^{(1)}(s_{12}, s_{13}, s_{23},\mu^2;\ep) + \Oe{0}\,.
\eeq

The infrared poles of the double virtual cross section cancel against the poles 
of the four integrated approximate cross sections in the sum (see \eqn{eq:sigmaNNLOm})
\beq
\dsig{NNLO}{3} \equiv
	\Big\{\dsig{VV}{3} 
	+ \int_2\Big[\dsiga{RR}{2}{5} 
	- \dsiga{RR}{12}{5}\Big] 
	+\int_1\Big[\dsiga{RV}{1}{4} 
	+ \Big(\int_1\dsiga{RR}{1}{5}\Big) \strut^{{\rm A}_{\scriptscriptstyle 1}} 
	\Big]\Big\} J_3\,.
\label{eq:dsigmaNNLO3}
\eeq
To indicate how this cancellation takes place, we use eqs.\ (\ref{eq:I1dsigRVA1}), 
(\ref{eq:I1dsigRA1_A1}), (\ref{eq:INTsigmaRRA12diff}) and (\ref{eq:INTsigmaRRA2diff}) 
to write the regularized double virtual cross section in the form
\beq
\bsp
\dsig{NNLO}{3} = 
	\Big\{\dsig{VV}{3} &+ 
		\dsig{B}{3} \otimes 
		\Big[\bI^{(0)}_{2}(\ep) - \bI^{(0)}_{12}(\ep) + \bI^{(1)}_{1}(\ep)
	+ \bI^{(0,0)}_{1,1}(\ep) 
	+ \frac{1}{2}\big\{\bI^{(0)}_{1}(\ep), \bI^{(0)}_{1}(\ep)\big\}\Big]
\\
	&+  \dsig{V}{3} \otimes \bI^{(0)}_{1}(\ep)
	\Big\} J_3\,.
\esp
\label{eq:VV3jet}
\eeq
The insertion operators appearing in \eqn{eq:VV3jet} above are given in terms of kinematic 
functions in eqs.\ (\ref{eq:I10}), (\ref{eq:I11}), (\ref{eq:I1100}), (\ref{eq:I12}) and 
(\ref{eq:I2}). We note that $\bI_{1}^{(0)}$ appears in \eqn{eq:VV3jet} multiplied by itself 
in the anti-commutator on the first line as well as by the virtual cross section on the second 
line. Since both $\bI_{1}^{(0)}$ and $\dsig{V}{3}$ contain up to $1/\ep^2$ poles, 
$\bI_{1}^{(0)}$  must be calculated to $\Oe{2}$ to correctly account for all finite parts 
in \eqn{eq:VV3jet}. In order to compute just the poles, it suffices to expand $\bI_{1}^{(0)}$ 
to $\Oe{}$ only, as in \appx{appx:I10}.
 
We have computed the pole parts of all insertion operators analytically, which turn out to be 
very lengthy expressions already at $\Oe{-2}$. (The reader can get an idea of the 
complexity by using the formulas in \appx{appx:I10} to compute the poles of 
$\big\{\bI^{(0)}_{1}(\ep), \bI^{(0)}_{1}(\ep)\big\}$.) However, the $\ep$-poles of the 
following combination of operators
\beq
\bom{J}_2 \equiv
  \bI_2^{(0)}
- \bI_{12}^{(0)}
+ \bI_{1}^{(1)}
+ \bI_{1,1}^{(0,0)}
+ \frac14\Big\{ \bI_{1}^{(0)}, \bI_{1}^{(0)} \Big\}
\label{eq:J2-def}
\eeq
form a remarkably simple expression:
\beq
\bsp
\bom{J}_2(\{p\}_3;\ep) = 
\frac{\as}{2\pi}\frac{\Sep}{\Fep} \left(\frac{\mu^2}{Q^2}\right)^\ep
\frac{1}{2\ep} \bigg[&
  \left(\beta_0 + 2\ep K - \ep^2 \beta_0 \frac{\pi^2}{4}\right)
  \bI_1^{(0)}(\{p\}_3;2 \ep)
\\&
- \beta_0\bI_1^{(0)}(\{p\}_3;\ep)
- \frac{\as}{2\pi}\frac{\Sep}{\Fep} \left(\frac{\mu^2}{Q^2}\right)^\ep
\Big(2 H_q(\Nf) + H_g(\Nf)\Big)
\bigg]
\\&
+ \Oe{0}\,.
\label{eq:J23j}
\esp
\eeq
It is easy to convince oneself that only the universal pole parts of 
the $\bI_1^{(0)}$ operator (given in \eqn{eq:I10poles} for general $m$) 
enter the computation of the poles of $\bom{J}_2$. Furthermore, looking at the 
explicit definition of $\bI_1^{(0)}$ in \eqn{eq:I10}, we see that the $\bom{J}_2$ 
operator in \eqn{eq:J23j} can be written by simply counting the radiating 
partons in the event (two quarks and one gluon in our example). This additive 
nature of $\bom{J}_2$, which is also valid for two-jet production, hints that in 
general
\beq
\bsp
\bom{J}_2(\{p\}_m;\ep) = 
\frac{\as}{2\pi}\frac{\Sep}{\Fep} \left(\frac{\mu^2}{Q^2}\right)^\ep
\frac{1}{2\ep} \bigg[&
  \left(\beta_0 + 2\ep K - \ep^2 \beta_0 \frac{\pi^2}{4}\right) 
  \bI_1^{(0)}(\{p\}_m;2 \ep)
\\&
- \beta_0\bI_1^{(0)}(\{p\}_m;\ep)
- \frac{\as}{2\pi}\frac{\Sep}{\Fep} \left(\frac{\mu^2}{Q^2}\right)^\ep
	\sum_{i=1}^m H_{f_i}(\Nf)
\bigg]
\\&
+ \Oe{0}
\,,
\label{eq:J2new}
\esp
\eeq
although presently we do not have a proof for the validity of this formula.
Using \eqns{eq:VV3jpoles}{eq:J23j} together with the explicit expressions
for $\bI_1^{(0)}$ in \appx{appx:I10}, it is not difficult to check explicitly 
that the combination
\beq
\bsp
\dsig{NNLO}{3} =
\bigg[\dsig{VV}{3}
	&+ \dsig{B}{3} \otimes \left(
  \bom{J}_2(\{p\}_3;\ep)
  + \frac14\Big\{
    \bI_{1}^{(0)}(\mom{}_{3};\ep),
    \bI_{1}^{(0)}(\mom{}_{3};\ep)
  \Big\} \right)
\\
	&+ \dsig{V}{3}\otimes \bI_{1}^{(0)}(\mom{}_{3};\ep)
\bigg] J_3\,,
\label{eq:dsigNNLO_3jet}
\esp
\eeq
is free of $\ep$-poles, although to perform the algebra for the $1/\ep^2$ and $1/\ep$
poles still requires some effort. Hence \eqn{eq:dsigNNLO_3jet} is finite in four 
dimensions and we can compute the regularized double virtual differential cross section 
for any infrared-safe observable numerically.

%%%
%%% Event shapes old and new
%%%
% ========== ========== ========== ========== ==========

\section{Event shapes old and new}
\label{sec:event}

The \colorfulNNLO\ method provides a robust subtraction scheme for 
computing NNLO corrections to processes with a colorless initial state 
(for the moment) and any number of final state jets, provided all necessary 
matrix elements are known. We have implemented the method in a general purpose, 
automated parton-level Monte Carlo code which can be used to compute any 
infrared-safe observable at NNLO accuracy in $e^+e^- \to 3$ jets. To demonstrate 
the validity of our code, we compute NNLO corrections to six standard event shape 
variables (thrust, heavy jet mass, total jet broadening, wide jet broadening, 
$C$-parameter and the two-to-three jet transition variable $y_{23}$ in the Durham 
algorithm) and compare our predictions to those available in the literature 
\cite{GehrmannDeRidder:2007hr,Weinzierl:2009ms}. We also present here for the first 
time the computation of jet cone energy fraction (JCEF) at NNLO accuracy. Predictions 
from \colorfulNNLO\ at this order in perturbation theory for oblateness and 
energy-energy correlation (EEC) were presented in ref.~\cite{DelDuca:2016csb}.

%%%
%%% Definition of event shapes
%%%

\subsection{Definition of event shapes}
\label{sec:eventshape-def}

Thrust~\cite{Brandt:1964sa,Farhi:1977sg} is defined as
\beq
T = \max_{\vec n} \left( \frac{\sum_i |\vec n\cdot \vec p_i|}{\sum_i |\vec p_i|} \right) \,,
\label{eq:thrust}
\eeq
where the three-vectors $\vec p_i$ denote the three-momenta of the partons and $\vec n$ 
defines the direction of the thrust axis, $\vec n_T$, by maximizing the sum on the 
right-hand side. For massless particles thrust is normalized by the center-of-mass energy, 
$\sum_i |\vec p_i| = Q$. In general $1/2\le T\le 1$, with $T=1/2$ for spherically 
symmetric events, and $T\to 1$ in the case of two back-to-back jets (the dijet limit).
For three-particle events, we have $2/3\le T\le 1$.

Heavy jet mass~\cite{Clavelli:1979md,Chandramohan:1980ry,Clavelli:1981yh}
is defined by dividing the event into two hemispheres, $H_L, H_R$, by a
plane orthogonal to an axis which can be chosen to be the thrust axis $\vec n_T$.
Then the hemisphere invariant mass is
\beq
\frac{M_i^2}{s} = \frac1{E_{\rm vis}^2} \bigg( \sum_{j\in H_i} p_j \bigg)^2\,, \qquad i=L, R\,,
\eeq
where $E_{\rm vis}$ is the total visible energy measured in the event,
which is equal to the center-of-mass energy in perturbation theory with
massless partons, $E_{\rm vis} = Q$.  The heavy jet mass is
\beq
\rho = \max \left( \frac{M_L^2}{s}, \frac{M_R^2}{s} \right)\,.
\label{eq:heavyjm}
\eeq
In the dijet limit, we find $\rho \to 0$. For three-particle events
we have $0\le \rho \le 1/3$.  At leading order in perturbation theory the 
distributions of heavy jet mass $\rho$ and $\tau \equiv 1-T$ are identical. 

Jet broadening~\cite{Rakow:1981qn,Catani:1992jc}, like heavy jet mass, is also defined 
through the two hemispheres $H_L, H_R$. First, hemisphere broadening is given by
\beq
B_i = \frac{ \sum_{j\in H_i} | \vec{p_j } \times \vec n_T | }{2 \sum_{j\in H_i} |\vec{p_j }| }\,, \qquad  i=L, R\,.
\label{eq:hembro}
\eeq
The total and wide jet broadening are then defined as
\beq
B_T = B_L + B_R
\qquad\mbox{and}\qquad
B_W = {\rm max} (B_L, B_R) \,.
\eeq
In the dijet limit, both $B_T$ and $B_W$ vanish, while for spherically
symmetric events $B_T = 2B_W = \pi/8$. For three-parton events we have 
$B_T, B_W \le 1/(2\sqrt{3}) \simeq 0.288$.

The $C$-parameter~\cite{Parisi:1978eg,Donoghue:1979vi} is defined
through the eigenvalues $\lambda_1, \lambda_2, \lambda_3$, of the
infrared-safe momentum tensor, 
\beq
\Theta^{\rho\sigma} = \frac1{\sum_i |\vec p_i|} \sum_i \frac{p_i^\rho p_i^\sigma}{|\vec p_i|}\,,
\qquad \rho,\sigma = 1, 2, 3\,,
\eeq
where $i$ runs over all final state particles.  As $\Theta$ is a
symmetric non-negative tensor with unit trace, the eigenvalues
$\lambda_i$ are real and non-negative, with $\sum_i\lambda_i = 1$.
Therefore, $0\le \lambda_i\le 1$, with $i=1, 2, 3$. The value of
the $C$-parameter is then defined as
\beq
C_{\rm par} = 
	3\left( \lambda_1 \lambda_2 + \lambda_2 \lambda_3 + \lambda_3 \lambda_1 \right)\,. 
\label{eq:Cpar}
\eeq
In the dijet limit the $C$-parameter vanishes, while for spherical events 
$C_{\rm par} = 1$, so $0\le C_{\rm par}\le 1$.  For events with three-partons 
in the final state we have $0\le C_{\rm par}\le 3/4$.  

Jet transition variables specify how an event changes from a $n$-jet to a $(n+1)$-jet 
configuration. For example, given a jet resolution parameter $y_{\rm cut}$, the 
two-to-three jet transition variable $y_{23}$\, 
\cite{Brown:1990nm,Catani:1991hj,Brown:1991hx,Bethke:1991wk}
is defined as the value of $y_{\rm cut}$ for which an event changes
from a two-jet to a three-jet configuration, within some jet algorithm.
Here we focus on the Durham algorithm~\cite{Bethke:1991wk}, which clusters
particles into jets by computing the variable,
\beq
y_{ij} = \frac{2 \min(E_i^2,E_j^2)(1-\cos\theta_{ij})}{E_{\rm vis}^2}\,,
\eeq
for each pair $(i, j)$ of particles. The pair with the lowest value of $y_{ij}$ 
is replaced by a pseudo-particle whose four-momentum is computed in the $E$ 
recombination scheme, i.e., it is simply the sum of the four-momenta of particles $i$ 
and $j$. This procedure is iterated until all pairs have $y_{ij} > y_{\rm cut}$ and 
the remaining pseudo-particles are the jets.

Finally, jet-cone energy fraction~\cite{Ohnishi:1994vp} is defined as the
energy deposited within a conical shell of the opening angle $\chi$
between a particle and the thrust axis $\vec n_T$, whose direction is
defined to point from the heavy jet mass hemisphere to the light jet
mass hemisphere, 
\beq
\frac{{\rm d}\Sigma_{\rm JCEF}}{{\rm d}\cos\chi} =
\sum_i \int \frac{E_i}{Q} \rd \sigma_{e^+e^-\to i + X}
\delta\bigg(\cos\chi - \frac{ \vec p_i \cdot \vec n_T }{|\vec p_i|}\bigg)\,.
\label{eq:JCEF}
\eeq
In principle $0^o \le\chi\le 180^o$, but hard gluon emissions typically contribute only 
to the region $90^o\le\chi\le 180^o$, which is plotted in the data~\cite{Abreu:2000ck}.

%%%
%%% Event shapes revisited
%%%

\subsection{Event shapes revisited}
\label{ssec:predictions}

\begin{figure}
\begin{center}
\includegraphics[width=0.60\textwidth]{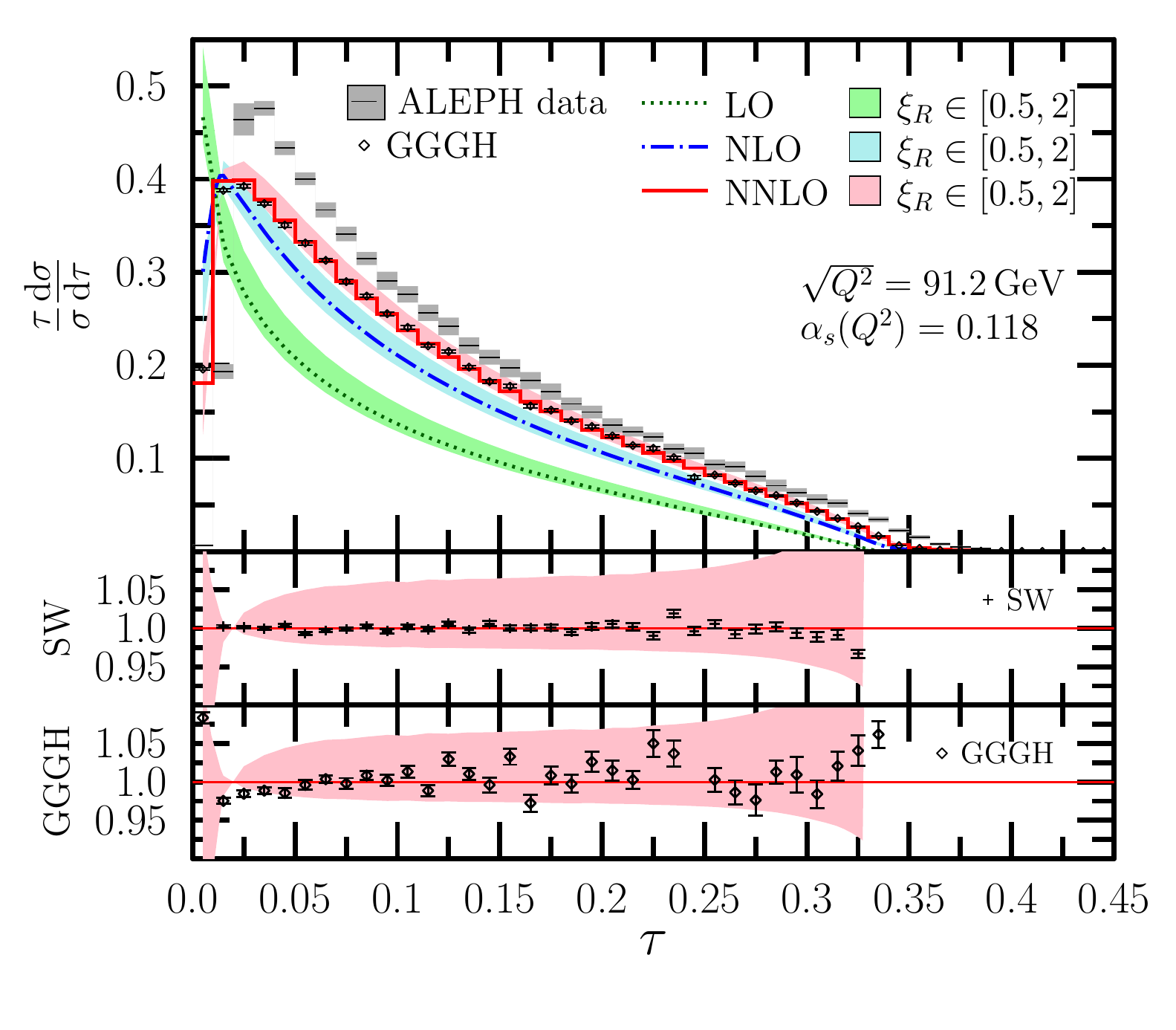}
\includegraphics[width=0.60\textwidth]{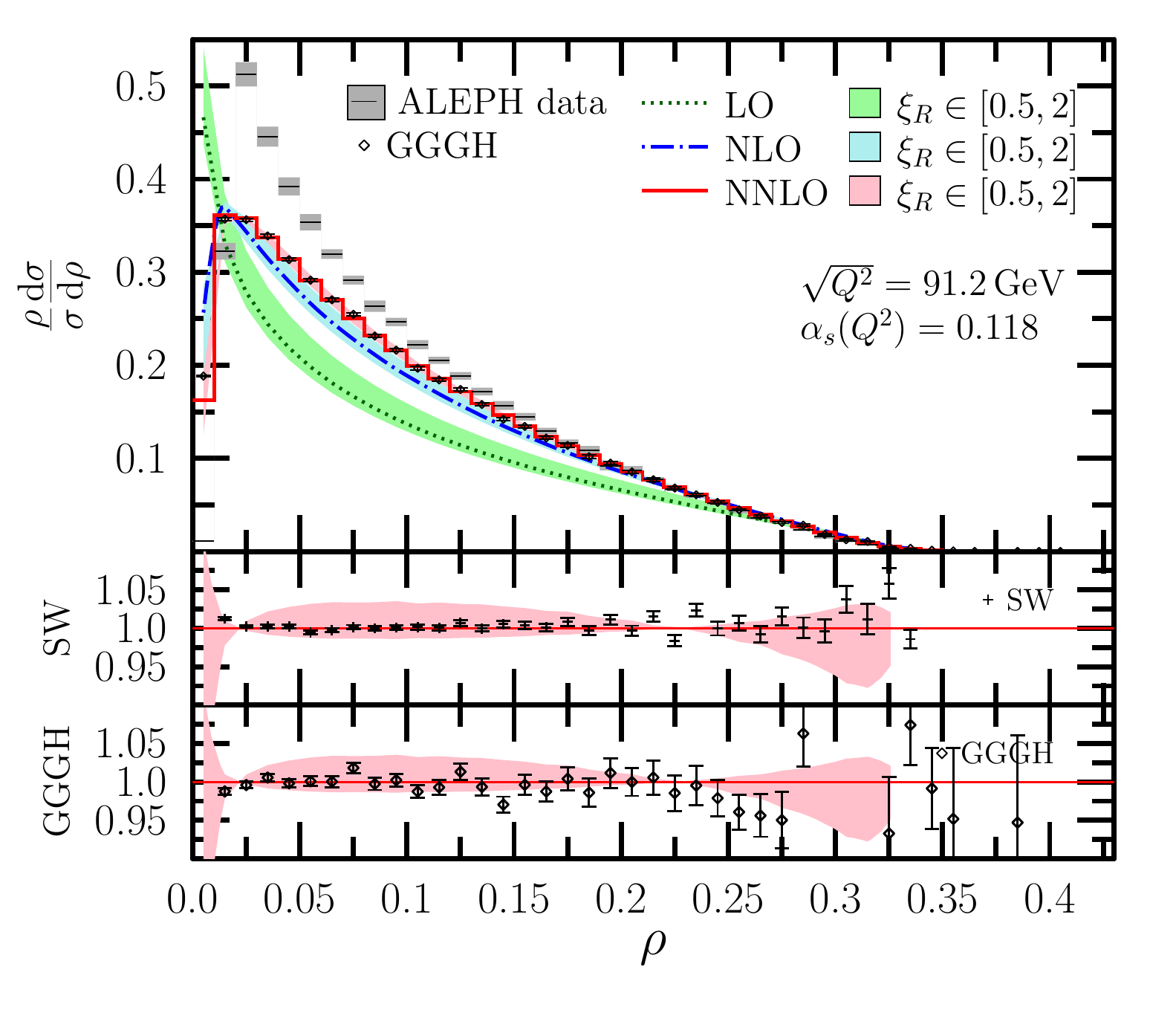}
\end{center}
\caption{\label{fig:physical-omT+rho} 
Perturbative predictions for the thrust ($\tau$) and heavy jet mass ($\rho$) distributions 
at LO, NLO and NNLO accuracy. The bands represent the renormalization scale uncertainty of 
our predictions corresponding to the range $\xi_R \in [0.5,2]$ around the central value of 
$\mu_0 = \sqrt{Q^2}$. The lower panels show the ratio of the (updated but unpublished -- 
see text) predictions of \cite{Weinzierl:2009ms} (SW) and \eerad~\cite{Ridder:2014wza} (GGGH) 
to \colorfulNNLO\ (this work). The bands on the lower panels show the relative scale uncertainty 
of our NNLO results.
}
\end{figure}

In this section we present the predictions of the \colorfulNNLO\ method for the 
event shapes considered also in \refrs{GehrmannDeRidder:2007hr,Weinzierl:2009ms}. 
To begin, we write the perturbative expansion of the differential distribution 
of an event shape observable $O$ at the default renormalization scale (not to be 
confused with the regularization scale of \sect{ssect:renorm}) $\mu_0 = \sqrt{Q^2}$ 
(the total center-of-mass energy) as
\beq
\frac{1}{\sigma_0}\frac{\rd \sigma}{\rd O} = 
  \frac{\as}{2\pi} A(O)
+ \left(\frac{\as}{2\pi}\right)^2 B(O)
+ \left(\frac{\as}{2\pi}\right)^3 C(O)
+ \Oa{4}\,,
\label{eq:O3expansion}
\eeq
where $\as = \as(\mu_0)$ and $\sigma_0$ is the leading-order perturbative 
prediction for the total cross section of the process $e^+e^-\to \mbox{hadrons}$.
The LO and NLO perturbative coefficients $A(O)$ and $B(O)$ for thrust, heavy jet mass, 
total and wide jet broadening, $C$-parameter and the jet transition variable $y_{23}$ 
in the Durham algorithm were computed a long time ago \cite{Kunszt:1989km}, while 
predictions for the NNLO coefficients $C(O)$ were presented in 
\cite{GehrmannDeRidder:2007hr,Weinzierl:2009ms}\footnote{Since these distributions 
have $1/O$ singularities, it is more convenient to present results for the quantities 
$O\, C(O)$ and this was done in \refrs{GehrmannDeRidder:2007hr,Weinzierl:2009ms} as 
well as in this paper in \figs{fig:physical-omT+rho}{fig:physical-Cpar+y23}.}. 
However, experiments measure 
the distributions normalized to the total hadronic cross section, $\sigma$, thus 
physical predictions should be normalized to that. At the default renormalization 
scale $\mu_0$, distributions normalized to the total hadronic cross section can be 
obtained from the expansion in \eqn{eq:O3expansion} above by multiplying with the 
inverse of
\beq
\frac{\sigma}{\sigma_0} =
1 + \frac{\as}{2\pi} A_{\rm t}
+ \left(\frac{\as}{2\pi}\right)^2 B_{\rm t}
+ \Oa{3}\,
\eeq
where \cite{Dine:1979qh,Chetyrkin:1979bj,Celmaster:1979xr}
\beq
A_{\rm t} = \frac32\CF
\quad\mbox{and}\quad
B_{\rm t} = \CF \left[
  \left(\frac{123}{8} - 11 \zeta_3\right) \CA
- \frac38 \CF + \left(4 \zeta_3 - \frac{11}{2}\right) \Nf \TR
\right]\,.
\eeq
The renormalization scale dependence of a three-jet event shape distribution 
normalized to the total hadronic cross section can be computed as
\beq
\frac{1}{\sigma} \frac{\rd \sigma(\mu)}{\rd O} = 
  \frac{\as(\mu)}{2\pi} \bar{A}(O;\mu)
+ \left(\frac{\as(\mu)}{2\pi}\right)^2 \bar{B}(O;\mu)
+ \left(\frac{\as(\mu)}{2\pi}\right)^3 \bar{C}(O;\mu) 
+ {\rm O(\as^4(\mu^2))}\,,
\label{eq:O3mudep}
\eeq
where
\beq
\bsp
\bar{A}(O;\mu) &= A(O)\,,
\\
\bar{B}(O;\mu) &= B(O) + \Big(\beta_0 \ln\xi_R - A_{\rm t}\Big) A(O)\,,
\\
\bar{C}(O;\mu) &= C(O) + \Big(2 \beta_0 \ln\xi_R - A_{\rm t}\Big) B(O)
+ \left(\frac12 \beta_1 \ln\xi_R + \beta_0^2 \ln^2\xi_R
+ A_{\rm t}^2 - B_{\rm t} \right) A(O)
\,,
\esp
\eeq
with $\xi_R \equiv \mu/\mu_0$.
Using three-loop running, the scale dependence of the strong coupling is given by
\beq
\frac{\as(\mu)}{2\pi} = \frac{2}{\beta_0 t}\left[1 
- \frac{\beta_1}{\beta_0^2 t} \ln t
+ \left(\frac{\beta_1}{\beta_0^2 t}\right)^2
\left(\ln^2 t - \ln t - 1 + \frac{\beta_0 \beta_2}{\beta_1^2}\right)
\right]
\,,
\eeq
with $t = \ln(\mu^2/\Lambda_{\rm QCD}^2)$. The first two coefficients 
in the expansion of the $\beta$ function,
\beq
\mu^2\frac{\rd}{\rd \mu^2} \frac{\as(\mu)}{4\pi} = 
- \left(\frac{\as(\mu)}{4\pi}\right)^2 \sum_{n = 0}^\infty \beta_n
  \left(\frac{\as(\mu)}{4\pi}\right)^n
\,,
\eeq
are presented in \eqn{eq:beta}, while \cite{Tarasov:1980au}
\beq
\beta_2 = \frac{2857}{54} \CA^3
- \left(\frac{1415}{27} \CA^2 + \frac{205}{9} \CA \CF - 2 \CF^2\right) \TR \Nf
+ \left(\frac{158}{27} \CA + \frac{44}{9} \CF\right) \TR^2 \Nf^2
\,.
\eeq
In order to compare to published predictions, we use $\as(m_Z) = 0.118$, 
corresponding to  $\Lambda_{\rm QCD} = 208$\,MeV.

We present physical predictions at the first three orders
in perturbation theory for the distributions of the six event shapes in
\figs{fig:physical-omT+rho}{fig:physical-Cpar+y23}. In the upper panels
we show our fixed-order predictions as well as those of the publicly available 
code \eerad\ \cite{Ridder:2014wza}\footnote{We are grateful to G.~Heinrich for 
providing the predictions of \eerad\ for us.}, together with the measured data by 
the ALEPH collaboration. We present our predictions at LO and NLO accuracy as smooth 
curves and as histogram at NNLO to represent the numbers of the numerical integration 
as precisely as possible. We observe a very good numerical convergence of our method 
at NNLO. The bands in the upper panel correspond to the variation of the renormalization 
scale in the range $\xi_R \in [0.5,2]$. In order to make the scale dependence 
at NNLO accuracy more visible, we show the relative scale uncertainty on the 
middle and bottom panels of each figure. It is remarkable that the relative scale 
dependence is below 5\,\% for most of the distributions in the ranges that are most 
relevant for measuring the strong coupling.  Nevertheless, there is still
a sizable difference between the NNLO predictions and the data for most of the 
distributions, which we attribute to parton shower (or resummation) and to 
hadronization effects. 

\begin{figure}
\begin{center}
\includegraphics[width=0.60\textwidth]{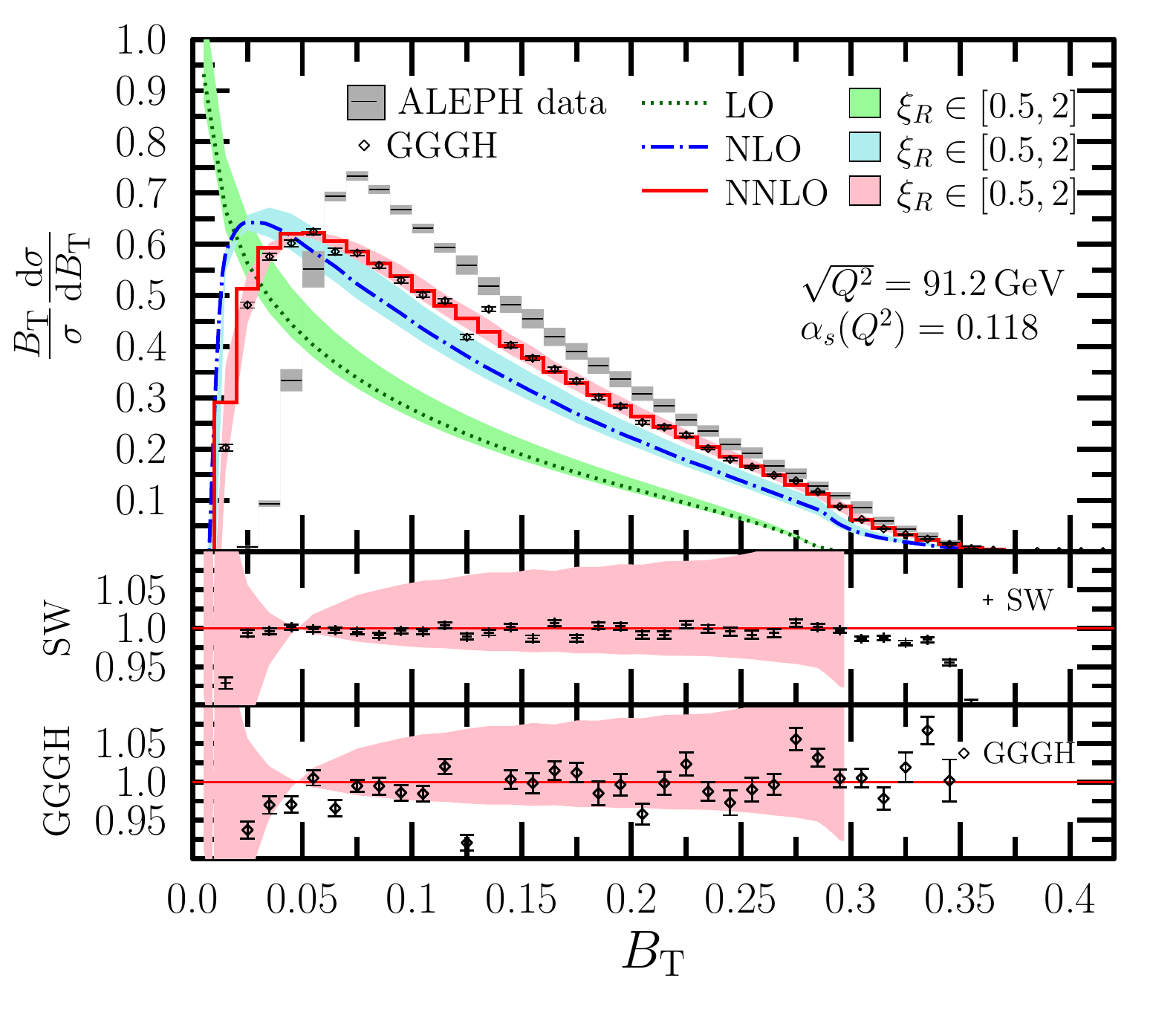}
\includegraphics[width=0.60\textwidth]{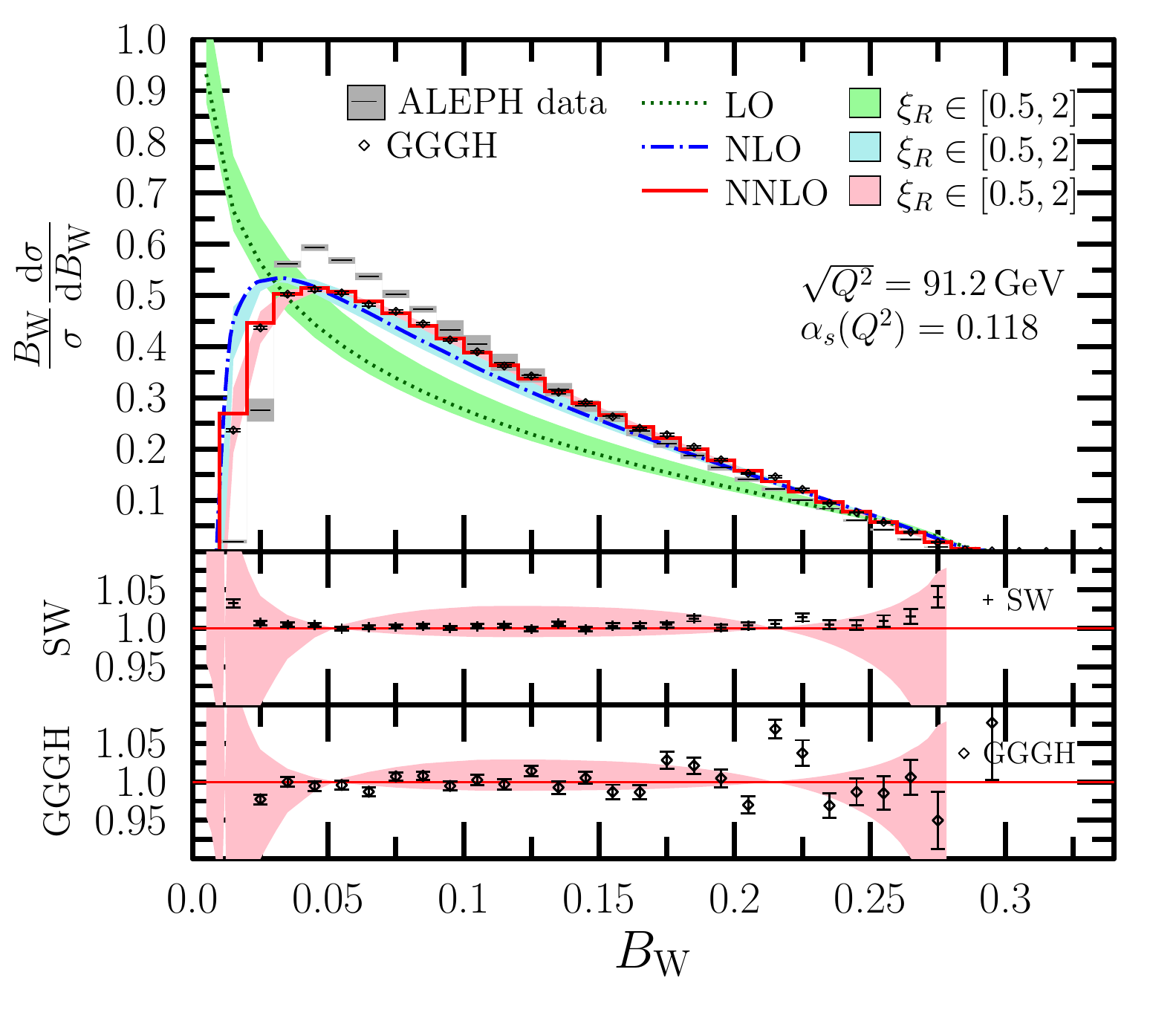}
\end{center}
\caption{\label{fig:physical-BT+BW} Same as \fig{fig:physical-omT+rho}
for total ($B_T$) and wide ($B_W$) jet broadening.}
\end{figure}

\begin{figure}
\begin{center}
\includegraphics[width=0.60\textwidth]{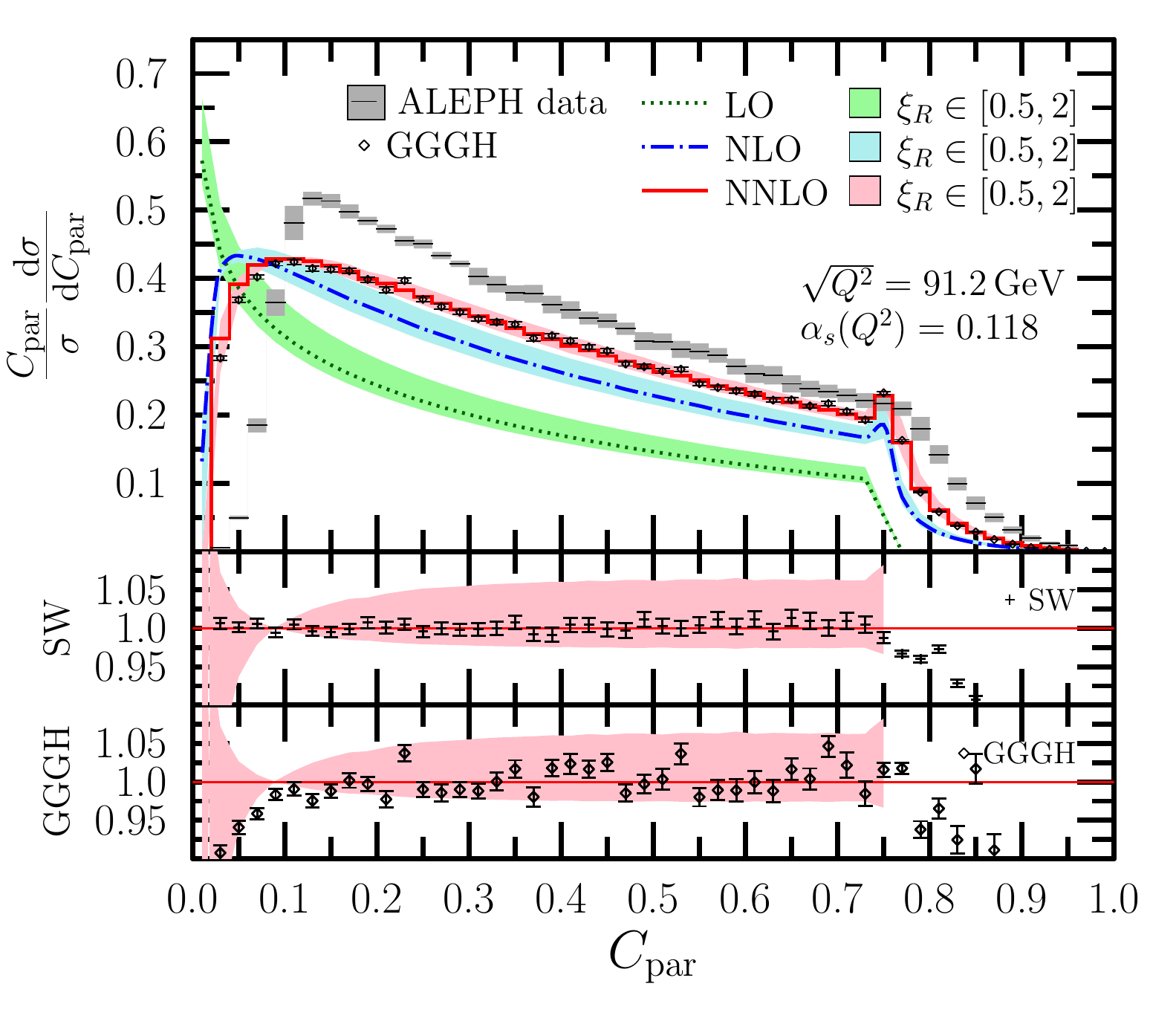}
\includegraphics[width=0.60\textwidth]{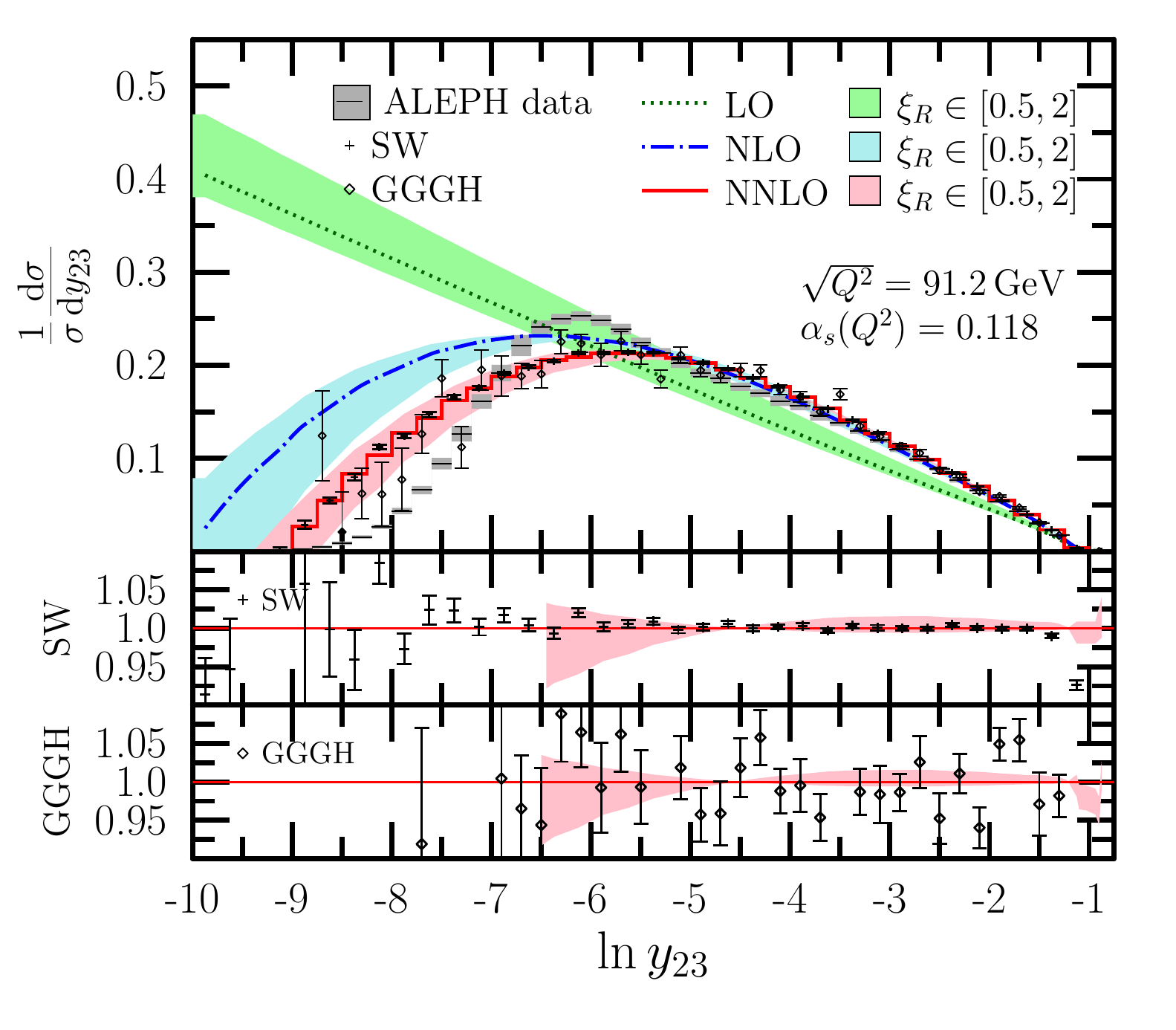}
\end{center}
\caption{\label{fig:physical-Cpar+y23} Same as \fig{fig:physical-omT+rho}
for the $C$-parameter ($C_{\rm par}$) and the two-to-three jet transition parameter
($y_{23}$) in the Durham clustering algorithm.}
\end{figure}

The dependence on the renormalization scale increases significantly 
beyond kinematical regions of three-parton contributions, for instance
for $\tau > 1/3$ or $C_{\rm par}>3/4$ and thus those are not shown on the 
ratio plots. At these three-parton kinematical limits large logarithms appear 
inside the physical region which have to be resummed similarly to the large
logarithms that appear for small values of the event shapes (at the
boundary of the physical region) \cite{Catani:1997xc}.

As mentioned above, predictions for these six event shapes were presented in 
\refrs{GehrmannDeRidder:2007hr,Weinzierl:2009ms}.
In order to quantify the level of agreement across the available perturbative predictions, 
we also show the ratio of the (updated but unpublished -- see below) results of 
\refr{Weinzierl:2009ms} (denoted by SW) and those of \eerad\ (denoted by GGGH) normalized 
to ours in the middle and bottom panels of each figure. Since the published predictions of 
\cite{Weinzierl:2009ms} are known to be affected by an issue with the phase space 
generation in the code used to compute those results \cite{Weinzierl:2010cw}, we have made 
comparisons to updated but unpublished results which where provided to us by 
S.~Weinzierl\footnote{We are grateful to S.~Weinzierl for providing us with these 
updated predictions.}. 
The general conclusion one may draw is that our predictions are in 
agreement with the updated predictions of SW except for very small and
large (beyond the kinematic limits at LO) values of the event shapes, 
up to the estimated statistical uncertainties. A qualitatively similar 
statement can be made about the comparison to the GGGH predictions, although 
the deviations from our results at small values of event shapes are in general 
more pronounced than for the SW predictions. This is especially apparent for 
the $C$-parameter distribution below $C_{\rm par} = 0.1$.

\begin{figure}
\includegraphics[width=0.45\textwidth]{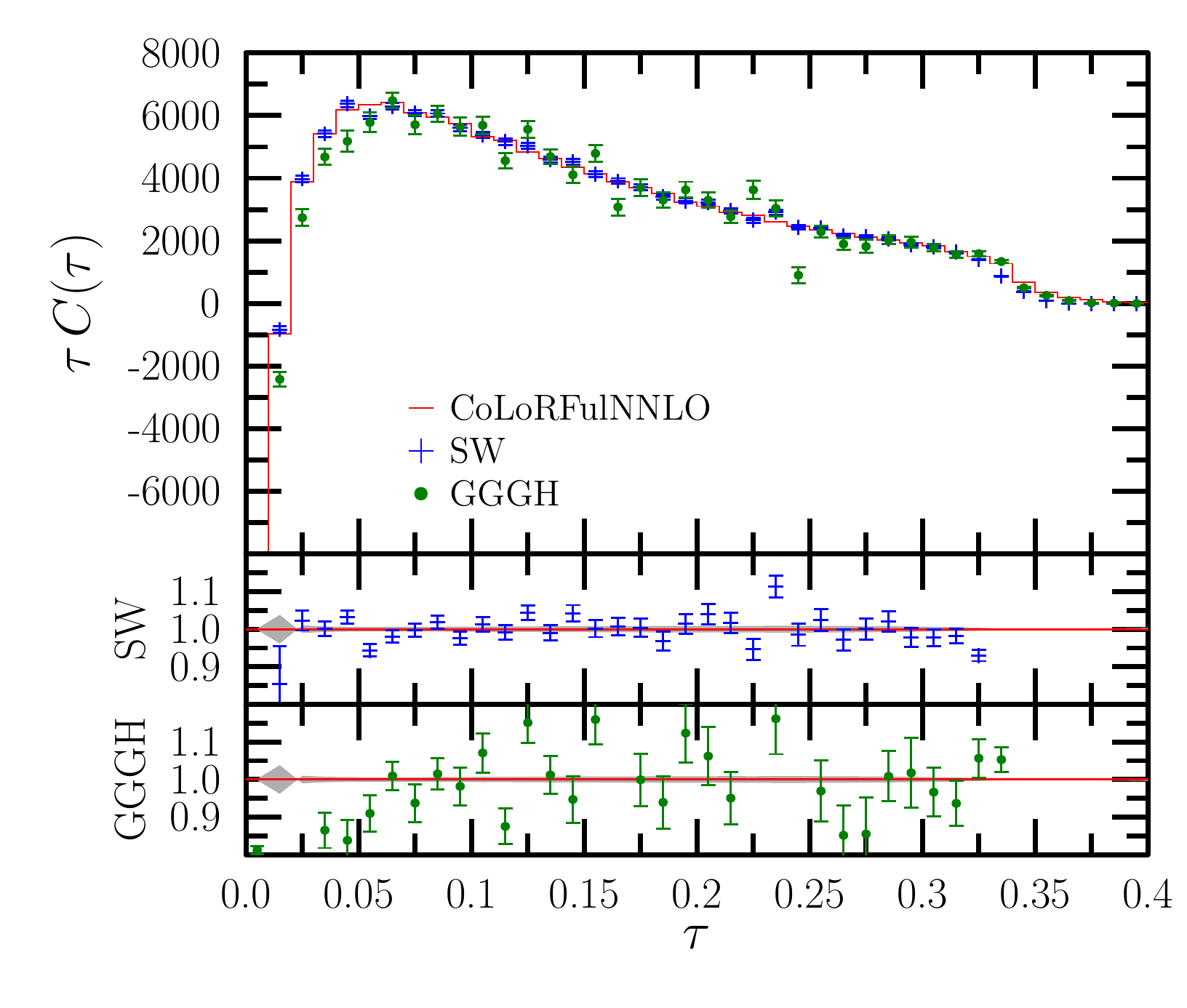}
\hspace{1em}
\includegraphics[width=0.45\textwidth]{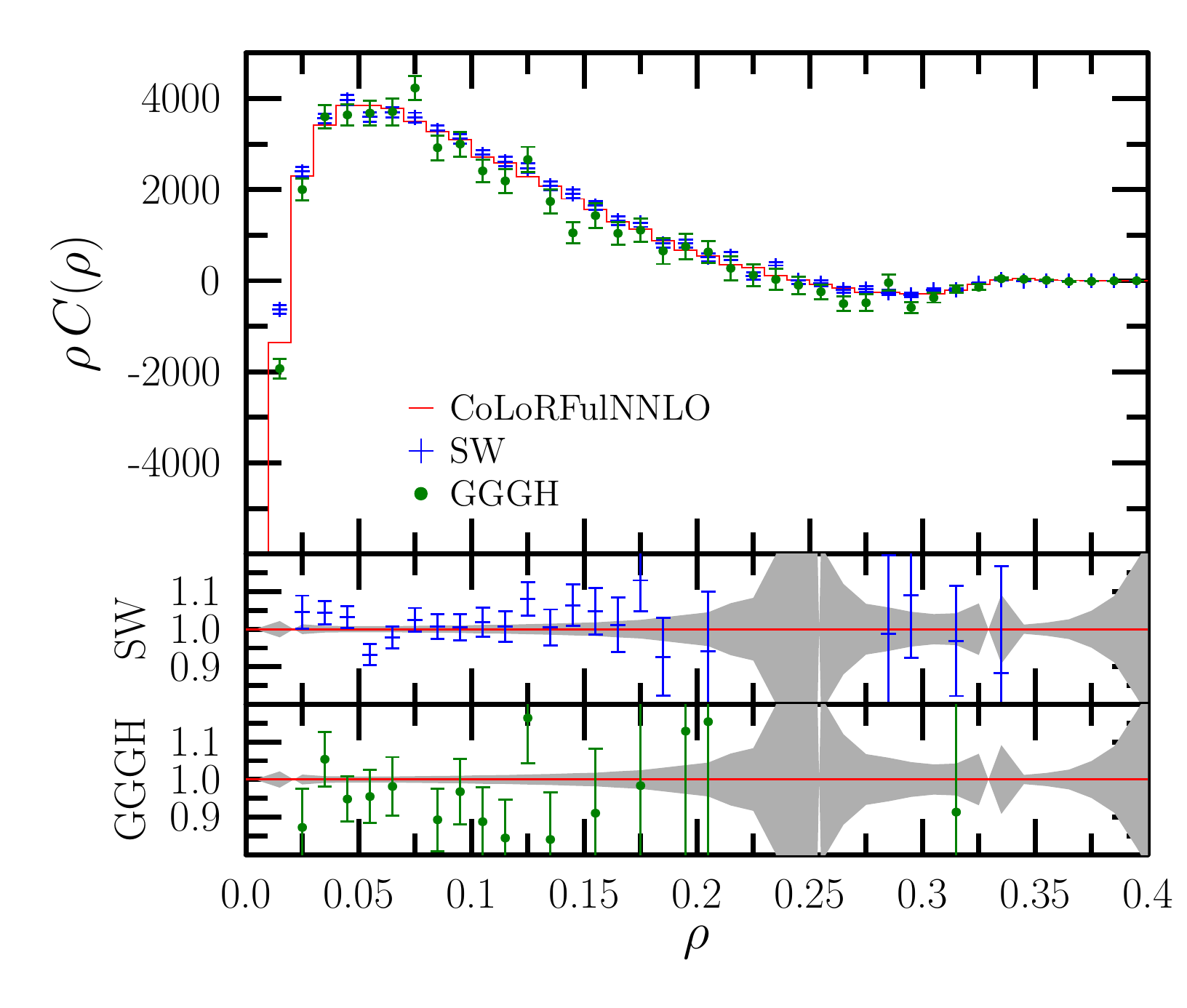}
\\
\includegraphics[width=0.45\textwidth]{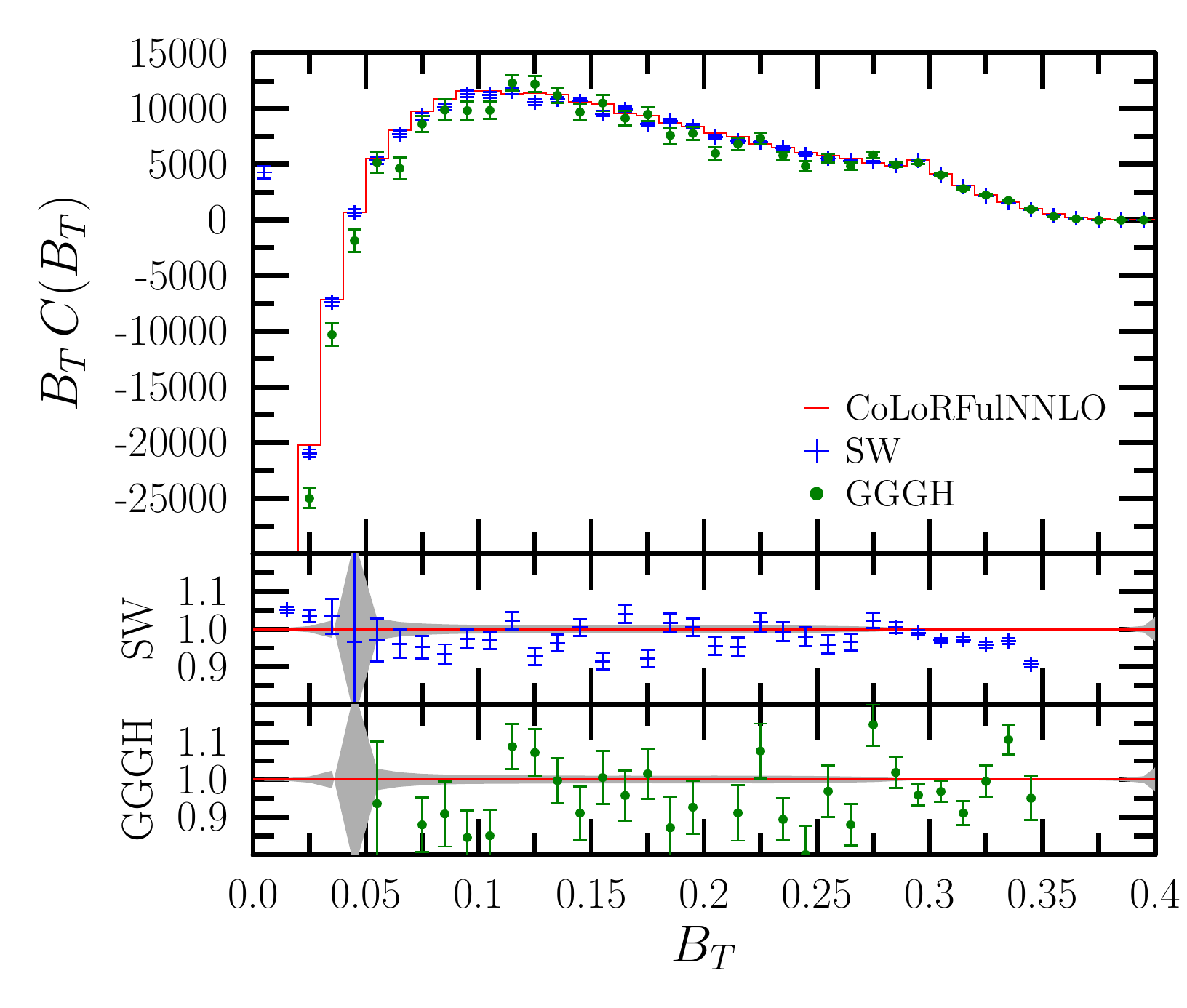}
\hspace{1em}
\includegraphics[width=0.45\textwidth]{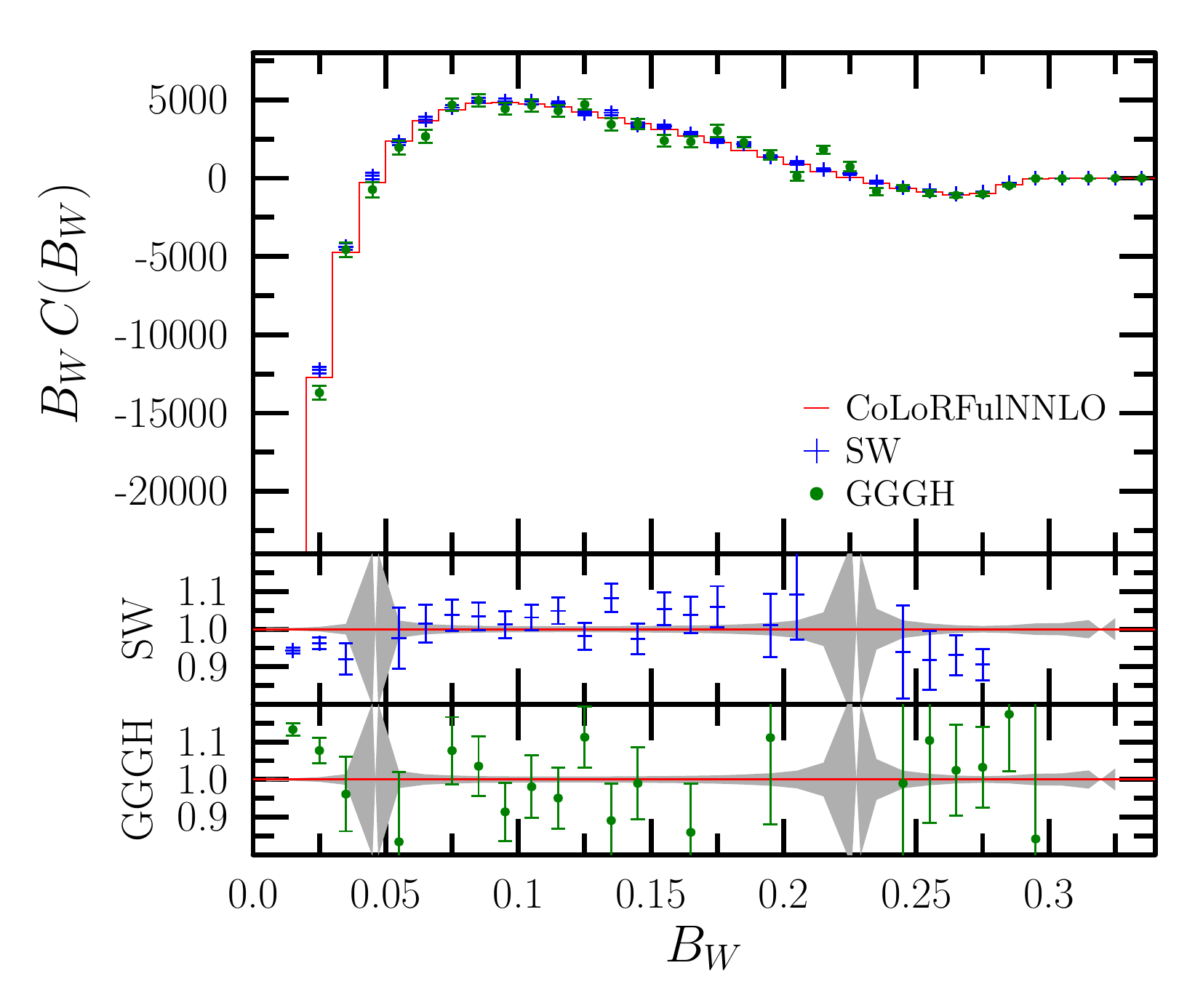}
\\
\includegraphics[width=0.45\textwidth]{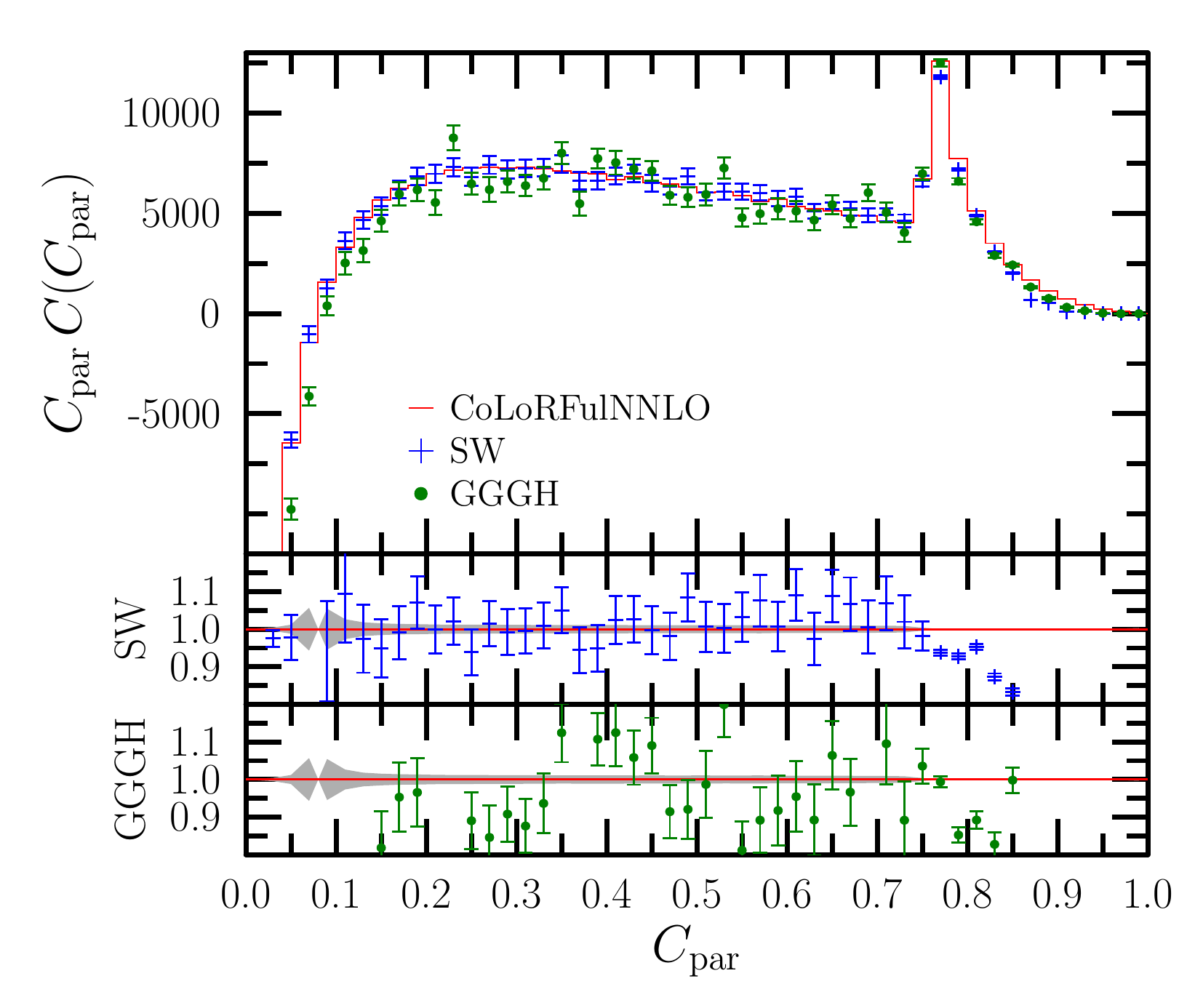}
\hspace{1em}
\includegraphics[width=0.45\textwidth]{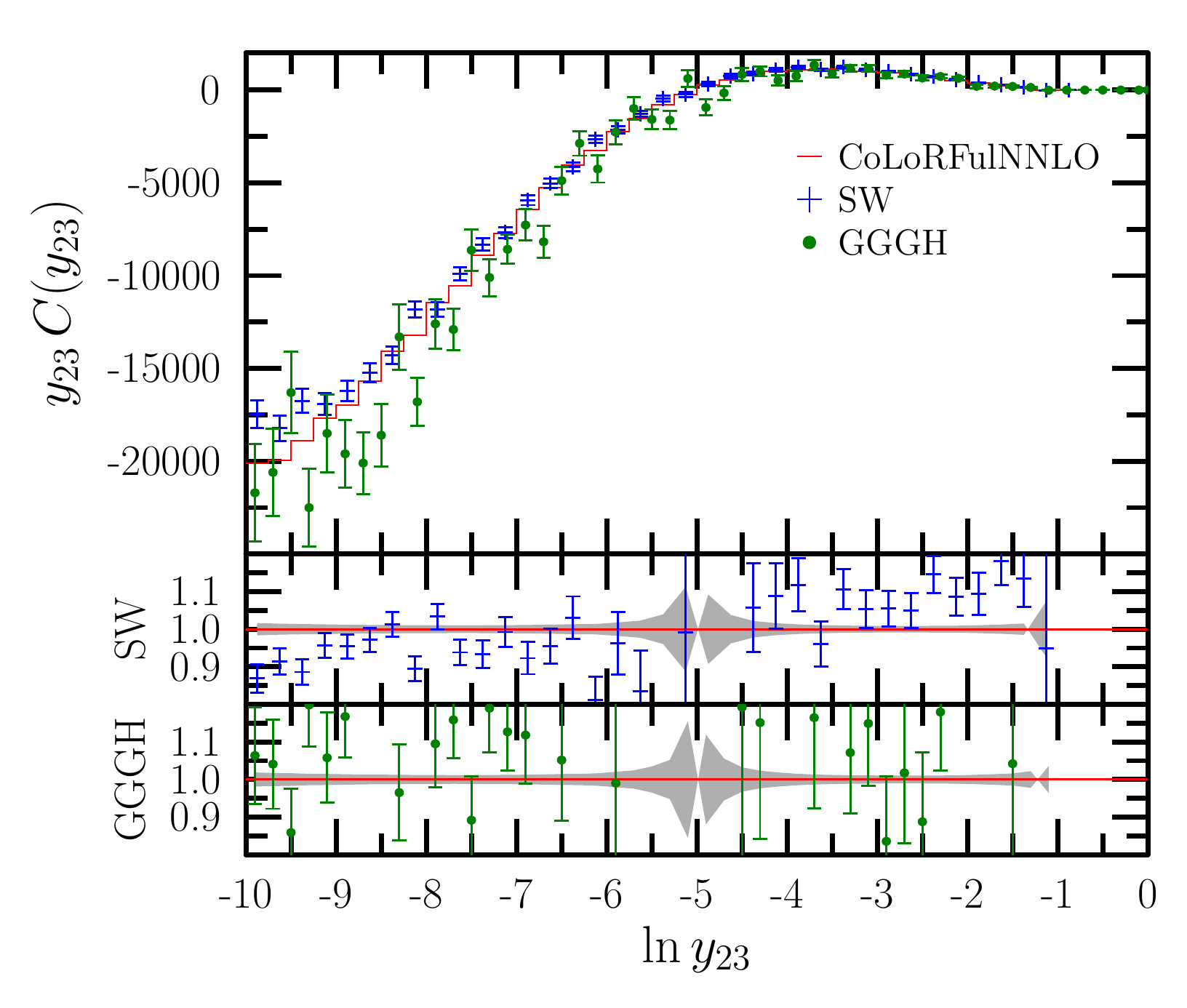}
\caption{\label{fig:Ccoeff-all} 
The $O\, C(O)$ coefficients of the thrust, heavy jet mass, 
total and wide jet broadening, $C$-parameter and two-to-three jet transition variable 
$y_{23}$ distributions. Lower panels show the ratio of the (updated but unpublished -- 
see text) predictions of \cite{Weinzierl:2009ms} (SW) and \eerad~\cite{Ridder:2014wza} (GGGH) 
to CoLoRFulNNLO (this work). The shaded bands on the lower panels represent the relative 
statistical uncertainties of our predictions due to Monte Carlo integrations.
}
\end{figure}

The level of agreement among the perturbative predictions can be seen 
better by looking at the NNLO coefficients directly, as shown in \fig{fig:Ccoeff-all}. 
In the figures the upper panels show the distributions of the NNLO coefficients 
$O\, C(O)$, while the middle and bottom panels once more present the ratios of the results 
of SW and those of GGGH normalized to ours. Again we observe a good numerical convergence 
of our method: the relative uncertainties of the Monte Carlo integrations are shown as 
shaded bands around the lines at one on the lower panels. 
The peaks which appear in the relative uncertainties are artifacts of the distributions 
changing sign with the absolute uncertainties remaining small. The scattered error bars 
represent the statistical uncertainties of the Monte Carlo integrations of the other two 
predictions.

Examining the plots in figure 4, we see that the agreement is generally quite good 
between the predictions of SW and \colorfulNNLO\ and reasonably good between GGGH and 
\colorfulNNLO. However the precise comparison to GGGH predictions is hampered by the 
somewhat large integration uncertainties and bin-to-bin fluctuations of those results.
Also, significant deviations among the three predictions are visible for small and
large values of the event shapes. For example for $\tau=1-T$ the differences 
between the \colorfulNNLO\ results and the other two computations grow up to a factor of two 
for $\tau > 1/3$. However, in this region, the contribution from three-particle final states 
vanishes and the thrust distribution is determined by a four-jet final state. Thus, $C(\tau)$ 
is determined by the NLO corrections to four-jet production, which have been known for a long 
time \cite{Signer:1996bf,Nagy:1997yn} and can also be computed with modern automated 
tools such as {\tt MadGraph5\_aMC@NLO} \cite{Alwall:2014hca}. We have checked that 
our predictions are in complete agreement with those of {\tt MadGraph5\_aMC@NLO}. 
The same is true for the tails of the other distributions beyond their respective 
kinematic limits. For small values of the event shapes we have checked that our
predictions agree with the resummed predictions obtained from SCET
\cite{Becher:2008cf,Chien:2010kc,Hoang:2014wka} expanded to \Oa{3}.

%%%
%%% Jet cone energy fraction
%%%

\subsection{Jet cone energy fraction}
\label{sec:predictions-jcef}

\begin{figure}
\begin{center}
\includegraphics[width=0.60\textwidth]{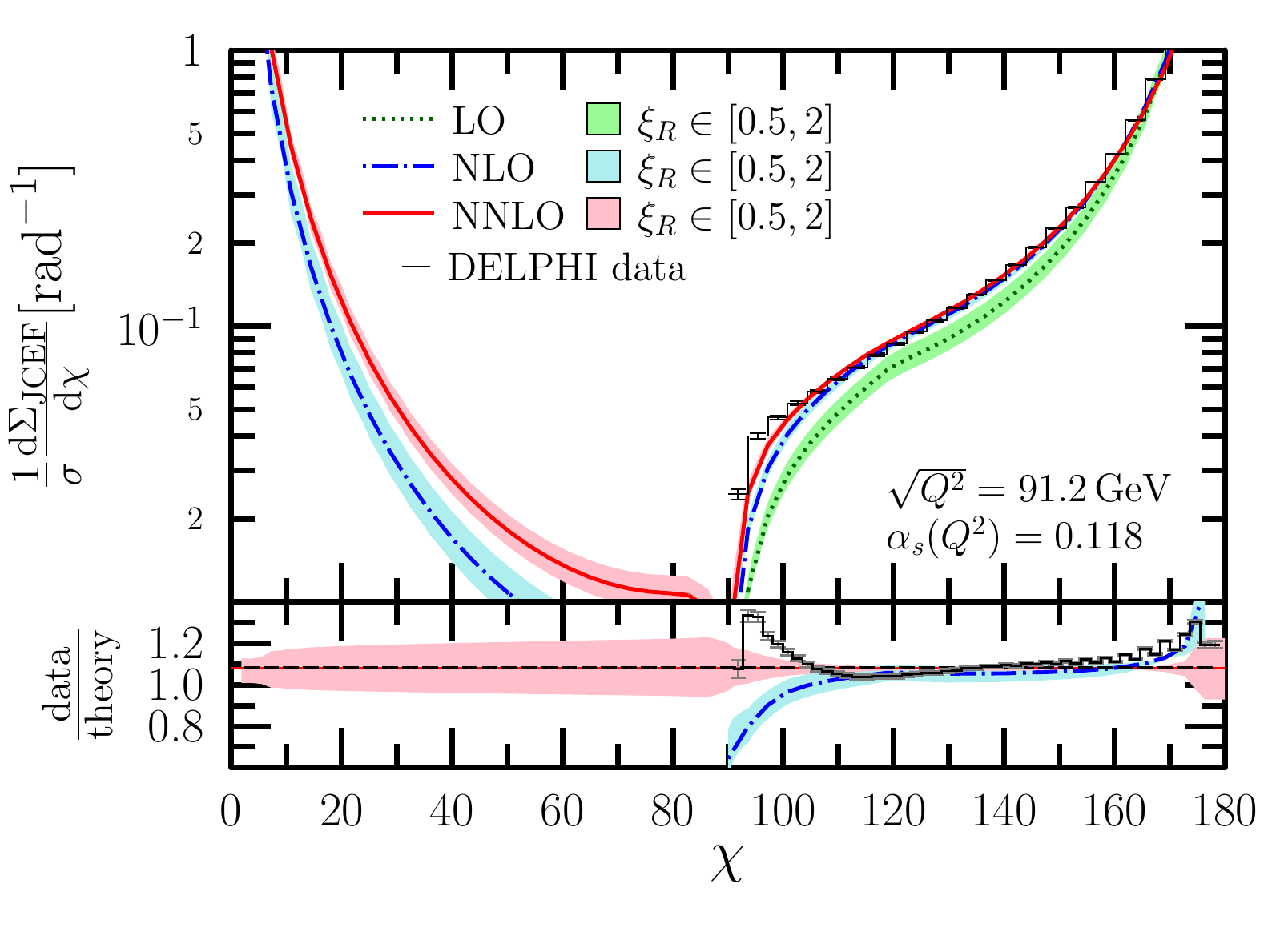}
\end{center}
\caption{\label{fig:physical-JCEF} The distribution of jet cone energy fraction (JCEF) 
at LO, NLO and NNLO accuracy in perturbative QCD. The bands represent the variation 
of the renormalization scale in the range $\xi_R = \mu/\mu_0 \in [0.5,2]$ around the 
central value of $\mu_0=\sqrt{Q^2}$. The lower panel shows the relative scale 
dependence at NNLO accuracy.}
\end{figure}

\begin{figure}
\begin{center}
\includegraphics[width=0.60\textwidth]{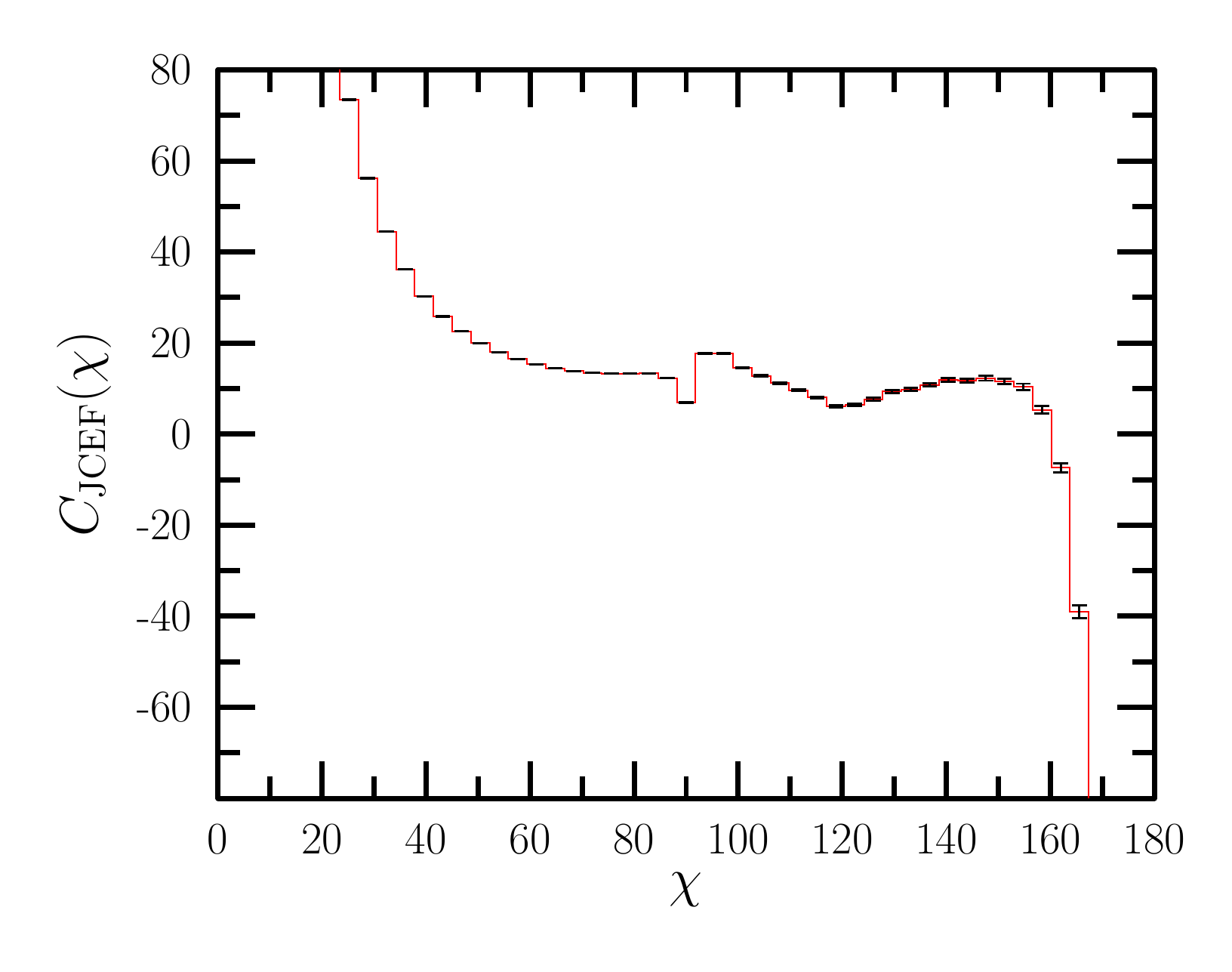}
\end{center}
\caption{\label{fig:Ccoeff-JCEF} The distribution of the NNLO coefficient for jet cone 
energy fraction (JCEF). The error bars represent the statistical uncertainty of the 
Monte Carlo integrations.}
\end{figure}

The jet cone energy fraction defined in \eqn{eq:JCEF} is a particularly simple 
and excellent observable for the determination of the strong coupling. The 
smallness of hadronization corrections, detector corrections as well as perturbative 
corrections allows a specially wide fit range to be used for the extraction of 
$\as$ \cite{Abreu:2000ck}. JCEF was computed at NLO accuracy 
for the first time in \refr{Ohnishi:1994vp}.
Here we present the first result for the JCEF distribution at NNLO accuracy in 
perturbative QCD for collider energy of $\sqrt{Q^2} = 91.2$\ GeV. 
In \fig{fig:physical-JCEF} we show physical predictions for JCEF, as well as the 
measured data by the DELPHI collaboration. As previously, the uncertainties due to 
the variation of the renormalization scale in the range $[0.5,2]$ times our default 
scale choice (the total center-of-mass energy) are shown as bands on the upper panel. 
We indicate the relative scale uncertainty at NNLO on the bottom panel. 
To better appreciate the impact of the NNLO corrections, we show in \fig{fig:Ccoeff-JCEF} 
the distribution of the NNLO coefficient $C(\chi)$ directly. Also for these distributions, 
we observe a good numerical convergence of our code.

%%%
%%% Conclusions and outlook
%%%
% ========== ========== ========== ========== ==========

\section{Conclusions and outlook}
\label{sec:conclusions}

In this paper we presented the \colorfulNNLO\ framework to compute higher order radiative 
corrections to jet cross sections in perturbative QCD. \colorfulNNLO\ is a completely 
local and fully differential subtraction scheme based on the known infrared factorization 
properties of QCD matrix elements in soft and collinear limits. Since the subtraction terms 
explicitly take all color and spin correlations into account, the regularized real emission 
terms (both double real and real-virtual) are well-defined and can be computed 
in four dimensions with whatever numerical procedure is deemed most convenient. We have shown 
analytically that explicit infrared $\ep$-poles coming from loop amplitudes cancel against 
the integrated forms of subtraction terms both in the real-virtual contribution (for any 
number of jets) and double virtual contribution (with up to three jets in the final state).

We have also reported on the computation of NNLO corrections to three-jet event shape 
observables in electron-positron collisions using \colorfulNNLO. We observe a very good 
numerical convergence of our method, which we attribute at least in part to the complete 
locality of the subtraction terms.

We compared both our physical predictions as well as the NNLO contribution only with 
similar predictions published earlier (in \refr{Ridder:2014wza} denoted by GGGH and 
in \refr{Weinzierl:2009ms} denoted by SW) for thrust, heavy jet mass, total and wide 
jet broadening, the C-parameter and the two-to-three jet transition variable $y_{23}$ 
in the Durham jet clustering algorithm. We find agreement with the updated (unpublished) 
predictions of SW within the statistical uncertainty of the numerical integrations except 
for very small and large values of the event shapes, beyond the kinematic limits at LO. 
The measured data in these regions are limited by statistics and so the phenomenological 
relevance of the differences is negligible. The same is true for the comparison to the 
physical predictions of GGGH, however the deviations from our results for small values 
of event shapes are generally more pronounced than for the predictions of SW. This is 
especially apparent for small values of the $C$-parameter, below $C_{\rm par} = 0.1$. 
When comparing our physical predictions at the NNLO accuracy to experimental data, we 
still find large differences. An important source of discrepancy is the neglected large 
logarithmic contributions which require all-order resummation. Work towards matching 
the fixed-order predictions to resummed ones is in progress.

Finally, we have shown for the first time perturbative predictions for jet cone energy 
fraction at NNLO, thereby providing a new observable from which the value of the strong 
coupling can be extracted at this accuracy. This observable has the remarkable property 
that the NNLO corrections are very small and the fairly good agreement between data and 
predictions already at NLO become only marginally better with the inclusion of the NNLO 
corrections. This stability of the perturbative predictions makes JCEF a good candidate 
for the extraction of the strong coupling.

We emphasize that our framework is not restricted to three-jet production, but it can be 
easily applied to study differential distributions for four- or more jet production once 
the necessary two-loop matrix elements become available. \colorfulNNLO\ is completely worked 
out for processes with colorless initial states at present. The inclusion of initial state 
radiation is a conceptually straightforward although substantial task and is work 
in progress.

%%%
%%% Acknowledgments
%%%
% ========== ========== ========== ========== ==========

\section*{Acknowledgments}
This research was supported by
the Hungarian Scientific Research Fund grant K-101482,
the SNF-SCOPES-JRP-2014 grant IZ73Z0\_152601,
the ERC Starting Grant ``MathAm'' and by
the National Science Foundation under Grant No. NSF PHY11-25915. 
AK acknowledges financial support from the
Post Doctoral Fellowship programme of the Hungarian Academy
of Sciences and the Research Funding Program ARISTEIA,
HOCTools (co-financed by the European Union (European Social Fund ESF).
We are grateful to U.~Aglietti, S.~Alioli, P.~Bolzoni, R.~Derco, 
S.-O.~Moch and D.~Tommasini for their contributions at intermediate
stages of this long project and to A.~Larkoski for his comment on the manuscript.

\section*{Note added}

After the completion of this manuscript, we were provided new
predictions of  \eerad\ by  G.~Heinrich.  In the present version of
this paper, we have made comparisons to these unpublished new predictions.

%%%
%%% Appendix
%%%

\appendix

%%%
%%% The $\bI_1^{(0)}$ insertion operator up to $\Oe{}$
%%%
% ========== ========== ========== ========== ==========

\section{The $\bI_1^{(0)}$ insertion operator up to $\Oe{}$}
\label{appx:I10}

We present the Laurent expansion of the kinematic functions that appear
in the $\bI_1^{(0)}(\{p\}_m;\ep)$ insertion operator in \eqn{eq:I10} up 
to and including \Oe{} terms. An expansion to this order is sufficient 
for demonstrating the cancellation of the $\ep$-poles at NNLO. We have also 
computed the $\Oe{2}$ coefficients of the expansions analytically, however 
they are quite lengthy and we do not display them here.

The $\IcC{1,i}{(0)}(x,\ep)$ are obtained as the following combination of terms
\bal
\IcC{1,q}{(0)}(x,\ep) &=
     [\IcC{ir}{(0)}]_{qg}(x,\ep)
     - [\IcC{ir}{}\IcS{r}{(0)}](\ep) \,,
\label{eq:C1q0}
\\
\IcC{1,g}{(0)}(x,\ep) &=
     \frac{1}{2}[\IcC{ir}{(0)}]_{gg}(x,\ep)  + \Nf [\IcC{ir}{(0)}]_{q\qb}(x,\ep)
     - [\IcC{ir}{}\IcS{r}{(0)}](\ep)\,,
\label{eq:C1g0}
\eal
where
\bal
& [\IcC{ir}{(0)}]_{qg}(x; \ep, \alpha_0 = 1, d_0 = 3 - 3\ep) = \frac{1}{\ep^2}+\left(\frac{3}{2}-2 \ln(x)\right)\frac1{\ep}
\nt\\& \quad
+ 2\left(1+\frac{1}{(1-x)^5}\right) \Li_{2}(1-x) -\frac{\pi ^2}{2} +2 \ln^{2}(x)
\nt\\& \quad
+\left(\frac{8}{3 (1-x)^5} - \frac{3}{2 (1-x)^4} - \frac{1}{3 (1-x)^3} + \frac{1}{3 (1-x)^2} + \frac{3}{2 (1-x)} -\frac{17}{3}\right) \ln(x)
\nt\\& \quad
+\frac{2}{3 (1-x)^4}-\frac{2}{3 (1-x)^3}-\frac{5}{12 (1-x)^2}+\frac{5}{24 (1-x)} +\frac{89}{24}
\nt\\& \quad
+ \ep\, \Bigg[
\left(10+\frac{26}{(1-x)^5}\right) \Li_{3}(1-x) + 20\left(1+\frac{1}{(1-x)^5}\right) \left(\Li_{3}(x) -\zeta_3\right)
\nt\\& \qquad
+ 8\left(1+\frac{1}{(1-x)^5}\right) \ln(x) \Li_{2}(1-x)
\nt\\& \qquad
-\frac{4 \ln^{3}(x)}{3} + 10\left(1+\frac{1}{(1-x)^5}\right) \ln(1-x) \ln^{2}(x) -\left(1+\frac{2}{(1-x)^5}\right) \pi ^2\ln(x)
\nt\\& \qquad
-\left(\frac{31}{3 (1-x)^5} +\frac{1}{2 (1-x)^4} +\frac{7}{3 (1-x)^3} +\frac{13}{3 (1-x)^2} +\frac{19}{2 (1-x)} -\frac{17}{3}\right) \Li_{2}(1-x)
\nt\\& \qquad
-\left(\frac{8}{3 (1-x)^5}-\frac{3}{2 (1-x)^4}-\frac{1}{3 (1-x)^3}+\frac{1}{3 (1-x)^2}+\frac{3}{2 (1-x)} -\frac{17}{3}\right) \ln^{2}(x)
\nt\\& \qquad
+\left(\frac{4}{3 (1-x)^4} +\frac{2}{3 (1-x)^3} +\frac{4}{9 (1-x)^2} +\frac{1}{3 (1-x)} +\frac{37}{36}\right) \pi^2
\nt\\& \qquad
+\left(\frac{1793}{72 (1-x)^5}-\frac{123}{8 (1-x)^4}-\frac{41}{36 (1-x)^3}+\frac{119}{36 (1-x)^2}+\frac{109}{8 (1-x)} -\frac{2327}{72}\right) \ln(x)
\nt\\& \qquad
+\frac{665}{72 (1-x)^4}-\frac{1069}{144 (1-x)^3}-\frac{343}{144 (1-x)^2}+\frac{737}{144 (1-x)}+\frac{353}{144}
\Bigg] 
+\Oe{2}\,,
\eal
\bal
& [\IcC{ir}{(0)}]_{q\qb}(x; \ep, \alpha_0 = 1, d_0 = 3 - 3\ep) = \frac{\TR}{\CA} \Bigg\{-\frac2{3 \ep}
\nt\\& \quad
+ \frac{2 }{3 } \left(1+\frac{1}{(1-x)^5}\right) \ln(x) - \frac{160}{3}  \left(\frac{2}{(2-x)^6}-\frac1{(2-x)^5}\right) \ln\left(\frac{x}{2}\right)
\nt\\& \quad
+\frac{2 }{3  (1-x)^4}+\frac{1}{3  (1-x)^3}+\frac{2 }{9  (1-x)^2}+\frac{1}{6  (1-x)}-\frac{5 }{2 }
\nt\\& \quad
-\frac{160 }{3  (2-x)^5}+\frac{40 }{3  (2-x)^4}+\frac{20 }{9  (2-x)^3}+\frac{5 }{9  (2-x)^2}+\frac{1}{6  (2-x)}
\nt\\& \quad
+ \ep\, \Bigg[
-\left(\frac{14 }{3  (1-x)^5}+\frac{2 }{3 }\right) \Li_{2}(1-x) - \frac{2 }{3 } \left(1+\frac1{(1-x)^5}\right) \ln^{2}(x) 
\nt\\& \qquad
+ \frac{160}{3}  \left(\frac{2}{(2-x)^6}-\frac1{(2-x)^5}\right)
\left( 8 \Li_{2}\left(1-\frac{x}{2}\right) - \Li_{2}(1-x) + \ln^{2}\left(\frac{x}2\right) \right)
\nt\\& \qquad
+\frac{7}{9 }\pi^2  - \frac{40}9 \left( \frac2{(2-x)^6} -\frac{1}{(2-x)^5}\right)  \pi ^2
+ \frac{992 }{9} \left(\frac{2}{(2-x)^6}-\frac{1}{(2-x)^5}\right) \ln(2)
\nt\\& \qquad
+\bigg(\frac{5 }{2  (1-x)^5}-\frac{1}{ (1-x)^4}-\frac{17 }{18  (1-x)^3}-\frac{11 }{6  (1-x)^2}-\frac{9 }{ (1-x)} + \frac{5 }{2 }
\nt\\& \qquad\quad
-\frac{1984 }{9  (2-x)^6}+\frac{1952 }{9  (2-x)^5}+\frac{80 }{9  (2-x)^3}+\frac{80 }{9  (2-x)^2}+\frac{8 }{ (2-x)}\bigg) \ln(x)
\nt\\& \qquad
+\frac{43 }{6  (1-x)^4}+\frac{25 }{12  (1-x)^3}+\frac{31 }{54  (1-x)^2}-\frac{10 }{9  (1-x)} -\frac{1399 }{108 }
\nt\\& \qquad
-\frac{4832 }{9  (2-x)^5}+\frac{1688 }{9  (2-x)^4}+\frac{644 }{27  (2-x)^3}+\frac{176 }{27  (2-x)^2}+\frac{25 }{9  (2-x)}
\Bigg] \Bigg\}
+\Oe{2}\,,
\eal
\bal
& [\IcC{ir}{(0)}]_{gg}(x; \ep, \alpha_0 = 1, d_0 = 3 - 3\ep) = \frac{2}{\ep^2}+ \left(\frac{11}{3}-4 \ln(x)\right)\frac{1}{\ep}
\nt\\& \quad
+4 \left(1+\frac{1}{(1-x)^5}\right) \Li_{2}(1-x) +4 \ln^{2}(x) -\pi ^2
+ \frac{160}{3}  \left(\frac{2}{(2-x)^6}-\frac1{(2-x)^5}\right) \ln\left(\frac{x}{2}\right)
\nt\\& \quad
+\left(\frac{14}{3 (1-x)^5}-\frac{3}{(1-x)^4}-\frac{2}{3 (1-x)^3}+\frac{2}{3 (1-x)^2}+\frac{3}{1-x}-12\right) \ln(x)
\nt\\& \quad
+\frac{2}{3 (1-x)^4}-\frac{5}{3 (1-x)^3}-\frac{19}{18 (1-x)^2}+\frac{1}{4 (1-x)} +\frac{37}{4}
\nt\\& \quad
+\frac{160}{3 (2-x)^5} -\frac{40}{3 (2-x)^4}-\frac{20}{9 (2-x)^3}-\frac{5}{9 (2-x)^2}-\frac{1}{6 (2-x)}
\nt\\& \quad
+ \ep \Bigg[ 40 \left(1+\frac{1}{(1-x)^5}\right) \left(\Li_{3}(x) - \zeta_3\right)
+\left(20+\frac{52}{(1-x)^5}\right) \Li_{3}(1-x) -\frac{8 \ln^{3}(x)}{3} 
\nt\\& \qquad
+ 16 \left(1+\frac{1}{(1-x)^5}\right) \ln(x) \Li_{2}(1-x) + 20\left(1+\frac{1}{(1-x)^5}\right) \ln(1-x) \ln^{2}(x)
\nt\\& \qquad
- 2\left(1+\frac{2}{(1-x)^5} \right) \pi^2 \ln(x)
\nt\\& \qquad
-\left(\frac{16}{(1-x)^5}+\frac{1}{(1-x)^4}+\frac{14}{3 (1-x)^3}+\frac{26}{3 (1-x)^2}+\frac{19}{1-x} - 12\right) \Li_{2}(1-x)
\nt\\& \qquad
- \frac{160}{3}  \left(\frac{2}{(2-x)^6}-\frac1{(2-x)^5}\right)
\left( 8 \Li_{2}\left(1-\frac{x}{2}\right) - \Li_{2}(1-x) + \ln^{2}\left(\frac{x}2\right) \right)
\nt\\& \qquad
-\left(\frac{14}{3 (1-x)^5}-\frac{3}{(1-x)^4}-\frac{2}{3 (1-x)^3}+\frac{2}{3 (1-x)^2}+\frac{3}{1-x} - 12\right)  \ln^{2}(x)
\nt\\& \qquad
+\left(\frac{8}{3 (1-x)^4} +\frac{4}{3 (1-x)^3} +\frac{8}{9 (1-x)^2} +\frac{2}{3 (1-x)} +\frac{23}{18} \right)\pi^2
\nt\\& \qquad
+ \frac{40}9 \left(\frac{2}{(2-x)^6}-\frac1{(2-x)^5}\right) \pi^2
- \frac{512}9 \left(\frac{2}{(2-x)^6}-\frac1{(2-x)^5}\right) \ln(2)
\nt\\& \qquad
+\bigg(\frac{1727}{36 (1-x)^5}-\frac{119}{4 (1-x)^4}-\frac{4}{3 (1-x)^3}+\frac{76}{9 (1-x)^2}+\frac{145}{4 (1-x)}- \frac{2393}{36}
\nt\\& \qquad\quad
+\frac{1024}{9 (2-x)^6}-\frac{1472}{9 (2-x)^5}-\frac{80}{9 (2-x)^3}-\frac{80}{9 (2-x)^2}-\frac{8}{2-x}\bigg) \ln(x)
\nt\\& \qquad
+\frac{431}{36 (1-x)^4}-\frac{1195}{72 (1-x)^3} -\frac{1105}{216 (1-x)^2}+\frac{829}{72 (1-x)} +\frac{3317}{216}
\nt\\& \qquad
+\frac{4352}{9 (2-x)^5} -\frac{1568}{9 (2-x)^4}-\frac{584}{27 (2-x)^3}-\frac{161}{27 (2-x)^2}-\frac{47}{18 (2-x)}\Bigg] 
+\Oe{2}
\,,
\eal
and
\beq[\IcC{ir}{}\IcS{r}{(0)}](\ep, 1, 3 - 3\ep) =  
\frac{1}{\ep^2} 
   + \frac{11}{3\ep}
   - \frac{7}{6} \pi ^2 + \frac{329}{18}
      + \left( -32 \zeta (3) -\frac{77}{18} \pi ^2 + \frac{9779}{108}\right) \ep
+\Oe{2}\,.
\eeq

The integrated soft kinematic function $\IcS{1}{(0),(i,k)}(Y,\ep)$ is simply given by 
\beq
\IcS{1}{(0),(i,k)}(Y,\ep) = [\IcS{1}{(0)}]^{(i,k)}(Y,\ep)\,,
\eeq
where
\bal
& [\IcS{r}{(0)}]^{(i,k)}(Y; \ep, y_0 = 1,  d_0' = 3 - 3\ep) = 
   - \frac{1}{\ep^2} 
   + \left( \ln (Y) -\frac{11}{3} \right) \frac{1}{\ep}
   \nt\\& \qquad
   -\Li_2(1-Y) -\frac1{2} \ln ^2(Y) + \frac{7}{6}\pi ^2 + \frac{11}{3} \ln (Y) -\frac{317}{18}
     \nt\\& \qquad
   + \bigg[
    -\Li_3(1-Y) -\Li_3(Y)+\frac1{6}\ln ^3(Y) -\frac{1}{2} \ln (1-Y) \ln ^2(Y)
    -\pi ^2 \ln (Y)+33 \zeta (3)
        \nt\\& \qquad
  + \frac{11}{6} \left( -2\Li_2(1-Y) - \ln ^2(Y) +  \frac{7}{3}\pi ^2\right) 
  + \frac{317 \ln (Y)}{18} -\frac{9299}{108}
\bigg] \ep 
+\Oe{2}\,.
\eal
 
%%%
%%% Asymptotic form of the $\bom{J}_2$ insertion operator
%%%
% ========== ========== ========== ========== ==========

\section{Asymptotic form of the $\bom{J}_2$ insertion operator}
\label{appx:I2}

We defined the $\bom{J}_2$ insertion operator in \eqn{eq:J2-def} and exhibited 
its pole structure for three hard partons in the final state in \eqn{eq:J23j}. 
That is sufficient to check the cancellation of the double virtual $\ep$-poles at 
NNLO. However, in the numerical integrations over the three-parton
phase space one also needs the finite part, which is rather lengthy and
in fact, we have only computed its asymptotic expansion for small kinematic 
invariants analytically. Here we record this expansion and comment on the remaining 
regular part, which we compute numerically.

For the Born process $e^+e^- \to q\qb g$ the $\bom{J}_2$ operator can be
written in the following explicit form:
\beq
\bsp
\bom{J}_{2}(\ep) =&
\left[\frac{\as}{2\pi} \frac{\Sep}{\Fep}
     \left(\frac{\mu^2}{Q^2}\right)^{\ep}\right]^2 \Bigg\{
\frac{1}{\ep^3} (\CA + 2 \CF)\Bigg( -\frac{3}{8}  \beta_0\Bigg)
\\ &+
\frac{1}{\ep^2} \Bigg[
  \frac14 \beta_0\bigg(\CA (\ln y_{13} + \ln y_{23}) - (\CA-2 \CF) \ln y_{12}\bigg)
- \frac{\pi^2}{24} \CA (\CA + 2 \CF)
\\ &\qquad
- \frac34 \CA^2 - \frac89 \CA \CF + \frac{17}{18} \CA \Nf \TR + \frac49 \CF \Nf \TR
- \frac29 \Nf^2 \TR^2\Bigg]
\\ &+
	\frac{1}{\ep} \Bigg[
- \frac{K}{2}
  \bigg(\CA (\ln y_{13} + \ln y_{23} ) - (\CA-2 \CF) \ln y_{12}\bigg)
\\ &\qquad
- \bigg(\frac{11}{36} \CA^2 + \frac12 \CF^2 - \frac19 \CA \Nf \TR\bigg) \pi^2
- \bigg(\frac14 \CA^2 + \frac{13}2 \CA \CF - 6 \CF^2\bigg) \zeta(3) 
\\ &\qquad
+ \frac{173}{54} \CA^2 + \frac{961}{216} \CA \CF - \frac{32}{27} \CA \Nf \TR + \frac38 \CF^2 -
\frac{46}{27} \CF \Nf \TR\Bigg]
\Bigg\}
\\&+ {\cal F}in\Big(\bom{J}_{2}(\ep)\Big) + \Oe{}\,,
\esp
\eeq
which then also defines the finite part ${\cal F}in\Big(\bom{J}_{2}(\ep)\Big)$ 
unambiguously. 
We decompose the finite part into an asymptotic piece which collects
all logarithmic contributions that become singular on the borders of the
three-parton phase space and a piece which is regular over the whole
phase space, i.e.~finite on the borders:
\beq
{\cal F}in\Big(\bom{J}_{2}(\ep)\Big) = 
	{\cal F}in\Big(\bom{J}_{2}^{\mathrm{asy}}(\ep)\Big) 
	+ {\cal F}in\Big(\bom{J}_{2}^{\mathrm{reg}}(\ep)\Big) \,.
\eeq
The asymptotic part can be written as
\beq
\bsp
{\cal F}in\Big(\bom{J}_{2}^{\mathrm{asy}}(\ep)\Big) &= 
	\left[\frac{\as}{2\pi} \frac{\Sep}{\Fep}
    \left(\frac{\mu^2}{Q^2}\right)^{\ep}\right]^2
\\& \times
	\sum_{i=1}^{3} \bigg[
		\IcC{2,i}{\mathrm{asy}}(x_i) \left(\bT_i^2\right)^2
		+ \sum_{\substack{k=1 \\ k\ne i}}^{3}
		\bigg(\IcS{2}{\mathrm{asy}}(Y_{ik})\, \CA 
			+ \IcSCS{2,i}{\mathrm{asy}}(x_i,Y_{ik})\, \bT_i^2\bigg) \bT_i \ldot \bT_k\bigg] \,.
\esp
\eeq
We stress that this form as well as the explicit expressions for the
asymptotic functions presented below are known to be appropriate only
for $e^+e^- \to 3$ jet production.  The analytic expressions for the 
asymptotic parts of the kinematic functions read as follows:
\bal
\IcC{2,q}{\mathrm{asy}}(x) =& 
%%%%%%%%%%%%%%%%
%%% ep^(0) part
%%%%%%%%%%%%%%%%
	- \frac{1}{9} \pi ^2 \ln (x)
	+ 12 \zeta (3) \ln (x)
	- \frac{27 \ln (x)}{4}
+ \frac{\CA}{\CF}
  \bigg[
	- \frac{22 \ln^3(x)}{9}
	- \frac{1}{3} \pi^2 \ln^2(x)
\nt \\ & 
	+ \frac{233 \ln^2(x)}{18} 
	- 3 \zeta (3) \ln (x) 
	- 4 \pi^2 \ln (2) \ln (x)
	+ \frac{7}{12} \pi ^2 \ln (x) 
	- \frac{2239 \ln (x)}{108}
  \bigg]
\nt \\ &
+ \Nf \frac{\TR}{\CF}
  \bigg[
	\frac{8 \ln^3(x)}{9}
	- \frac{38}{9} \ln^2(x)
	+ \frac{\pi^2}{3} \ln (x)
	+ \frac{137 \ln (x)}{27}
  \bigg]\,,
\eal
\bal
\IcC{2,g}{\mathrm{asy}}(x) =& 
%%%%%%%%%%%%%%%%
%%% ep^(0) part
%%%%%%%%%%%%%%%%
	- \frac{22}{9} \ln^3(x) 
	- \frac{1}{3} \pi^2 \ln^2(x) 
	+ \frac{85 \ln^2(x)}{6} 
	+ 9 \zeta (3) \ln (x) 
	- 4 \pi^2 \ln (2) \ln(x)
	+ \frac{37}{36} \pi^2 \ln (x)
\nt \\ &
	- \frac{884 \ln (x)}{27}
+ \Nf \frac{\TR}{\CA}
  \bigg[
	\frac{8 \ln ^3(x)}{9}
	- \frac{64 \ln ^2(x)}{9}
	+ \frac{5}{9} \pi ^2 \ln(x)
	+ \frac{107 \ln (x)}{9}
  \bigg]
\nt \\ &
+ \Nf \frac{\CF \TR}{\CA^2}
  \bigg[
	- \frac{4}{3} \pi ^2 \ln (x)
	+ \frac{14 \ln (x)}{3}
  \bigg]
+ \Nf^2 \frac{\TR^2}{\CA^2}
  \bigg[
	\frac{8 \ln ^2(x)}{9}
	- \frac{52 \ln (x)}{27}
  \bigg]\,,
\eal
\bal
\IcS{2}{\mathrm{asy}}(Y) &= 
%%%%%%%%%%%%%%%%
%%% ep^(0) part
%%%%%%%%%%%%%%%%
	\frac{11 \ln^3(Y)}{18}
	+ \frac{13}{6} \pi^2 \ln^2(Y)
	- \frac{244 \ln^2(Y)}{9}
	+ 17 \zeta (3) \ln (Y)
	- 8 \ln^2(2) \ln (Y)
	- \frac{13}{4} \pi ^2 \ln (Y)
\nt \\ &
	+ \frac{217}{9}\ln (2) \ln (Y)
	+ \frac{131167 \ln(Y)}{2700}
+ \frac{\CF}{\CA}
  \bigg[
	- \frac{4}{3} \pi^2 \ln^2(Y)
	+ \frac{98 \ln^2(Y)}{9}
	- 16 \zeta (3) \ln (Y)
\nt \\ &
	+ \frac{26}{3} \pi^2 \ln (Y)
	- \frac{5099 \ln (Y)}{75}
  \bigg]
+ \Nf \frac{\TR}{\CA}
  \bigg[
	-\frac{2}{9} \ln^3(Y)
	+ \frac{20 \ln^2(Y)}{9}
	+ 16 \ln^2(2) \ln(Y)
\nt \\ &
	- \frac{29}{9} \pi^2 \ln(Y)
	- \frac{434}{9} \ln (2) \ln (Y)
	+ \frac{4279 \ln (Y)}{135}
  \bigg]\,,
\eal
\bal
\IcSCS{2,q}{\mathrm{asy}}(x,Y) &= 
%%%%%%%%%%%%%%%%
%%% ep^(0) part
%%%%%%%%%%%%%%%%
	8\bigg(\frac{1}{(1-x)^5}+1\bigg) 
	\bigg(
		2\Li_3(1-x) \ln (Y)
		+ \Li_2(1-x) \ln (x) \ln (Y)
\nt \\ &
		+ 2 \Li_3(x) \ln (Y)
		+ \ln (1-x) \ln ^2(x) \ln (Y)
		- \frac{\pi^2}{6} \ln (x) \ln (Y)
		- 2\zeta (3) \ln (Y)
	\bigg)
\nt \\ &
	- 8 \bigg(
		\Li_3(1-x) \ln (Y)
		- \Li_2(1-Y) \ln (x) \ln (Y)
		- \Li_3(1-Y) \ln (x)
		- 2 \Li_3(Y) \ln (x)
\nt \\ &		
		-\ln (x) \ln (1-Y) \ln ^2(Y)	
		+ \frac{\pi^2}{3} \ln (x) \ln (Y)
		- \frac{\pi^2}{6} \ln^2(Y)
		- 4\zeta (3) \ln (Y)
	\bigg)
\nt \\ &		
	- \frac{4}{3} \bigg(
		2 \Li_2(1-Y) \ln (x)
		+ \ln (x) \ln ^2(Y)
	\bigg)
	- \bigg(\frac{38}{3 (1-x)^5}-\frac{2}{(1-x)^4}+\frac{2}{(1-x)^3}
\nt \\ &		
		+\frac{4}{(1-x)^2}
		+\frac{8}{1-x}-\frac{16}{3}\bigg) \Li_2(1-x) \ln (Y)
	- \frac{107 \ln ^2(Y)}{9}
	+ \bigg(\frac{4}{3 (1-x)^4}
\nt \\ &	
	+\frac{2}{3(1-x)^3}+\frac{4}{9 (1-x)^2}
		+\frac{1}{3 (1-x)}-4\bigg) \pi^2 \ln (Y)
	+ \bigg(\frac{157}{18 (1-x)^5}-\frac{5}{3 (1-x)^4}
\nt \\ &
	-\frac{53}{18 (1-x)^3}
		+\frac{16}{9 (1-x)^2}+\frac{35}{3 (1-x)}
		-\frac{104}{9}\bigg) \ln (x) \ln (Y)
	+ \frac{4}{1-Y} \ln (x) \ln (Y)
\nt \\ &
	+ \bigg(\frac{97}{18 (1-x)^4}-\frac{77}{36(1-x)^3}-\frac{17}{9 (1-x)^2}
		+\frac{145}{36 (1-x)}+\frac{9247}{450}\bigg) \ln (Y)
\nt \\ &
+ \frac{\CA}{\CF}
  \bigg[
	- \frac{4}{3} \pi^2 \ln^2(Y)
	+ \frac{35 \ln^2(Y)}{3}
  	- 16 \zeta (3) \ln (Y)
	+ \frac{22}{9} \pi ^2 \ln (Y)
	+ 8 \ln^2(2) \ln (Y)
\nt \\ &
	- \frac{224}{9} \ln (2) \ln (Y)
	+ \frac{9629 \ln (Y)}{1350}
  \bigg]
+ \Nf \frac{\TR}{\CF}
  \bigg[
  	\frac{4 \ln^2(Y)}{9}
	+ \frac{28}{9} \pi^2 \ln (Y)
	- 16 \ln^2(2) \ln (Y)
\nt \\ &
	+ \frac{448}{9} \ln (2) \ln (Y)
	- \frac{7474 \ln (Y)}{135}
  \bigg]\,,
\eal
and
\bal
\IcSCS{2,g}{\mathrm{asy}}(x,Y) &= 
%%%%%%%%%%%%%%%%
%%% ep^(0) part
%%%%%%%%%%%%%%%%
	8\bigg(\frac{1}{(1-x)^5}+1\bigg)
	\bigg(
		2 \Li_3(1-x) \ln (Y)
		+ \Li_2(1-x) \ln (x) \ln (Y)
\nt \\ &   
  		+ 2 \Li_3(x) \ln (Y)
		+ \ln (1-x) \ln^2(x) \ln (Y)
		- \frac{\pi^2}{6} \ln (x) \ln (Y)
		-2 \zeta (3) \ln (Y)
	\bigg)
\nt \\ &   
  -8 \bigg(
		\Li_3(1-x) \ln (Y)
  		- \Li_2(1-Y) \ln (x) \ln (Y)
		- \Li_3(1-Y) \ln (x)
\nt \\ &   
		- 2 \Li_3(Y) \ln (x)
		- \ln (x) \ln (1-Y) \ln^2(Y)
		+ \frac{\pi^2}{3} \ln (x) \ln (Y)
		- 2 \zeta (3) \ln (Y)
	\bigg)
\nt \\ & 
	- \frac{4}{3} \bigg(2 \Li_2(1-Y) \ln (x)+\ln (x) \ln ^2(Y)\bigg)
	- \bigg(\frac{34}{3 (1-x)^5}-\frac{2}{(1-x)^4}
\nt \\ & 	
	+\frac{2}{(1-x)^3}
		+\frac{4}{(1-x)^2}+\frac{8}{1-x}-\frac{16}{3}\bigg) \Li_2(1-x) \ln (Y)
	- \frac{320}{3} \bigg(\frac{2}{(2-x)^6}
\nt \\ & 	
	-\frac{1}{(2-x)^5}\bigg) 
		\Li_2\bigg(1-\frac{x}{2}\bigg) \ln (Y)
	+ \bigg(\frac{4}{3 (1-x)^4}+\frac{2}{3 (1-x)^3}
	+\frac{4}{9 (1-x)^2}
\nt \\ & 	
		+\frac{1}{3 (1-x)}\bigg) \pi ^2 \ln (Y)
	+ \bigg(\frac{163}{18 (1-x)^5}-\frac{7}{3 (1-x)^4}-\frac{55}{18 (1-x)^3}
		+\frac{16}{9 (1-x)^2}
\nt \\ & 		
		+\frac{35}{3 (1-x)}-\frac{100}{9}\bigg) \ln (x) \ln (Y)
	- \bigg(\frac{96}{(2-x)^6}-\frac{272}{3 (2-x)^5}+\frac{56}{3 (2-x)^4}
\nt \\ & 
		+\frac{4}{3 (2-x)^3}\bigg) \ln \bigg(\frac{x}{2}\bigg) \ln (Y)	
	+ \frac{4}{1-Y} \ln (x) \ln (Y)
	+ \bigg(\frac{79}{18 (1-x)^4}-\frac{107}{36 (1-x)^3}
\nt \\ & 	
	-\frac{64}{27 (1-x)^2}
		+\frac{15}{4 (1-x)}\bigg) \ln (Y)
	+ \bigg(\frac{176}{3 (2-x)^5}-\frac{20}{3 (2-x)^4}-\frac{154}{27 (2-x)^3}
\nt \\ & 
		-\frac{37}{27 (2-x)^2}-\frac{5}{18 (2-x)}\bigg) \ln (Y)
\nt \\ & 
  + \Nf \frac{\TR}{\CA}
  \bigg[
  	\frac{640}{3}\bigg(\frac{2}{(2-x)^6}-\frac{1}{(2-x)^5}\bigg) 
		\Li_2\bigg(1-\frac{x}{2}\bigg) \ln (Y)
\nt \\ & 
	- \frac{8}{3 (1-x)^5} \Li_2(1-x) \ln (Y)
	- \bigg(\frac{2}{3 (1-x)^5}-\frac{4}{3 (1-x)^4}-\frac{2}{9 (1-x)^3}
\nt \\ & 
		+\frac{8}{9}\bigg) \ln (x) \ln (Y)
	+ \bigg(\frac{192}{(2-x)^6}-\frac{544}{3 (2-x)^5}+\frac{112}{3 (2-x)^4}
\nt \\ & 
		+\frac{8}{3 (2-x)^3}\bigg) \ln \bigg(\frac{x}{2}\bigg) \ln (Y)
	+ \bigg(\frac{2}{(1-x)^4}+\frac{5}{3 (1-x)^3}+\frac{26}{27 (1-x)^2}
\nt \\ & 
		+\frac{5}{9 (1-x)}\bigg) \ln (Y)
	- \bigg(\frac{352}{3 (2-x)^5}-\frac{40}{3 (2-x)^4}-\frac{308}{27 (2-x)^3}
\nt \\ & 
		-\frac{74}{27 (2-x)^2}-\frac{5}{9 (2-x)}\bigg) \ln (Y)		
  \bigg]\,.
\eal

We do not have analytic expressions for the regular part. However, computing 
this piece numerically on a grid over the three-parton phase space, we find 
that it is in fact flat across the whole phase space (within the uncertainty 
of the numerical integrations). Hence it can be described by a single number 
whose numerical value we find to be
\beq
{\cal F}in\Big(\bom{J}_{2}^{\mathrm{reg}}(\ep)\Big) = 	
	\left[\frac{\as}{2\pi} \frac{\Sep}{\Fep}
    \left(\frac{\mu^2}{Q^2}\right)^{\ep}\right]^2 \bigg(-650\bigg)
\eeq
for $\Nc=3$, $\TR=1/2$ and $\Nf=5$.

%%%
%%% Bibliography
%%%
% ========== ========== ========== ========== ==========

%\bibliographystyle{utphys}
%\bibliography{ee3jets_paper}

\providecommand{\href}[2]{#2}\begingroup\raggedright\endgroup

%%%
%%% End document
%%%

\end{document}